# Evaluating and Minimizing Distributed Cavity Phase Errors in Atomic Clocks

*Ruoxin Li and Kurt Gibble, Department of Physics, The Pennsylvania State University,*

*University Park, PA 16802, USA*

**ABSTRACT**

We perform 3D finite element calculations of the fields in microwave cavities and analyze the distributed cavity phase errors of atomic clocks that they produce. The fields of cylindrical cavities are treated as an azimuthal Fourier series. Each of the lowest components produces clock errors with unique characteristics that must be assessed to establish a clock's accuracy. We describe the errors and how to evaluate them. We prove that sharp structures in the cavity do not produce large frequency errors, even at moderately high powers, provided the atomic density varies slowly. We model the amplitude and phase imbalances of the feeds. For larger couplings, these can lead to increased phase errors. We show that phase imbalances produce a novel distributed cavity phase error that depends on the cavity detuning. We also design improved cavities by optimizing the geometry and tuning the mode spectrum so that there are negligible phase variations, allowing this source of systematic error to be dramatically reduced.

## 1. INTRODUCTION AND BACKGROUND

Atomics clocks may have a systematic error of their frequency due to phase variations distributed throughout the microwave cavity(ies). Phase variations exist because the microwave losses in the cavity walls create small travelling waves. These phase variations couple with the atomic motion to produce a first order Doppler shift. The resulting distributed cavity phase (DCP) error is the leading contribution to the uncertainty of some of the world's best atomic clocks [1-4]. Surprisingly, there are no clock uncertainty evaluations that quantitatively treat all of the relevant DCP errors. Here we discuss the general behaviors of DCP errors, describe methods to set limits on their uncertainties, and demonstrate techniques to design cavities with negligible DCP errors.

There is a 35 year history of work on spatial phase variations in cavities and DCP errors, dating back to atomic beam clocks [5]. Nearly all atomic fountain clocks use $TE_{011}$ cylindrical cavities, which have small losses and a high Q [4]. In 1996, Khursheed *et al.* solved the two-dimensional finite element problem for the phase variations, assuming no longitudinal variation [6]. Jefferts *et al.* showed that the transverse phase gradients could be made much smaller by using more azimuthally distributed feeds [7]. For this cavity, it was claimed that the maximum phase deviation was ±0.34 µrad for 4 feeds, which leads to negligible DCP errors. However, the first three dimensional solutions followed shortly thereafter and showed a longitudinal variation of more than 100 µrad, as well as the radial phase variation for a cavity with 2 feeds [8]. Unfortunately, full three dimensional finite element solutions require vastly more computer resources than two dimensional solutions so it was not possible to see features that require dense meshing or to perform many calculations [9]. In 2004, we showed that the three dimensional fields could be expressed as a short Fourier series in cos(mϕ) of two dimensional problems in ρ and z for cylindrical coordinates $\vec{r}=(\rho,\phi,z)$ [10]. This led to an analytic solution for cavities with no endcap holes, giving insight as well as an often accurate description of the phase in the center of real cavities. This solution showed that endcap losses produce the large longitudinal phase variation and a radial phase variation of order 5 μrad, which cannot be reduced by adding more azimuthally distributed feeds. The Fourier series in $\phi$ also enabled efficient finite element solutions for cavities with holes in the endcaps using dense meshes. We presented the azimuthally symmetric (m=0) fields in [10] and a key result was that the fields and the losses are very large near the apertures of the endcap holes - the phase of the field approaches $\pi/4$ [10].

The existence of large phase shifts in the cavity is inconsistent with the premise of the early accuracy evaluations of fountain clocks [4], as well as some more recent evaluations [11,12]. In 2005, two groups reported calculations of the power dependence of DCP errors due to azimuthally symmetric (m=0) longitudinal phase variations [13,14]. We used our finite element solutions from [10]. Ref [14] instead used a phenomenological field that is not a solution of Maxwell's equations. The m=0 phase variation that they analyzed is of order 100 times smaller than the dominant m=0 DCP error, has a dramatically different shape as a function of power, and arises from different physical effects [13]. It is important to note that this power dependent template is used by a number of groups to evaluate their clock's DCP uncertainty [3,12], including the NIST and INRIM clocks which report DCP uncertainties smaller than $3\times10^{-17}$ [15,16], an order of magnitude smaller than those for other clocks. Another recent treatment used an incomplete mode expansion that does not satisfy Maxwell's equations at 9.2 GHz or the boundary conditions [2]. A number of DCP uncertainty estimates no doubt include some element of intuition, with a goal of being conservative. Except for the NIST and INRIM clocks, the DCP uncertainties for other clocks are $3\times10^{-16}$ or larger. Most of the evaluations of small uncertainties do not address the large phase shifts near the apertures. None address the large longitudinal phase gradients from the 3D calculations even though the first work showed that the longitudinal phase gradients produce local Doppler shifts of $10^{-13}$ [8] and others have warned that longitudinal imbalances give a high sensitivity to first order Doppler shifts [17].

Here we show that DCP errors can be evaluated using full 3D calculations of the cavity fields by separating the fields and DCP errors into Fourier components in $\phi$. Despite the long history of DCP errors, no models of DCP errors, for either fountains or beam clocks, have reasonably described the experimental measurements. Recent work comparing the models of this work to preliminary measurements appears promising [18,19]. With experimental verification, these models should allow lower DCP uncertainties to be demonstrated.

In the next section, we review the basic features of the fields and the phase variations that follow from Maxwell's equations [10]. In Section 3, Schrödinger's equation gives the atomic response to the phase variations and we describe our calculations of DCP errors. In Section 4, we use the analytic phase variations from [10] to demonstrate the general features of DCP errors, both their origin and behavior, especially as a function of microwave power. Section 5 describes more general behaviors of DCP errors, those related to amplitude and phase imbalances of the feeds, particularly for strongly coupled cavities. In Section 6, we discuss real cavities, first proving that sharp features, where the phase shifts are as large as $\pi/4$, produce negligible DCP errors. We describe our finite element calculations, especially for m≥1. We then describe the DCP error for the cavities used in current fountain clocks due to each Fourier component of the fields. Finally, we show techniques to design cavities with small phase variations that produce negligible DCP errors, even at elevated power.

## 2. FIELDS OF TE$_{011}$ CYLINDRICAL MICROWAVE CAVITIES

We begin by describing the electromagnetic field of TE$_{011}$ cylindrical microwave cavities. We highlight key results from Ref. [10] and here we explicitly give the frequency dependence of the fields and phase variations. We write the fields as a superposition of a large standing wave $\vec{E}_0(\vec{r})$ and $\vec{H}_0(\vec{r})$ and a small standing wave $\vec{f}(\vec{r})$ and $\vec{g}(\vec{r})$, that describe the losses [6,10].

$$\vec{H}(\vec{r}) = \vec{H}_0(\vec{r}) + \left[\alpha(\Delta\omega) + i\right]\vec{g}(\vec{r})$$
$$\vec{E}(\vec{r}) = i\vec{E}_0(\vec{r}) - \left[1 - \alpha(\Delta\omega)i\right]\vec{f}(\vec{r})$$
(1)

All fields have an $e^{-i\omega t}$ time dependence and are in general complex. The form of (1) also allows the fields $\vec{E}_0(\vec{r}), \vec{H}_0(\vec{r}), \vec{f}(\vec{r})$, and $\vec{g}(\vec{r})$ to be real [10].

In (1), $\alpha(\Delta\omega)$ describes the dominant frequency dependence of the fields for detunings that are comparable to the cavity linewidth. The cavity fullwidth is $\Gamma_F=\omega/Q$ where the loaded Q's typically range from 2,000 to essentially the maximum of 30,000 for copper. With the cavity driven at the atomic transition frequency, $\omega=2\pi\times9.1926$ GHz for cesium, we can imagine tuning the cavity by shrinking or enlarging it, for example, as in temperature tuning. When the cavity is too large such that $\Delta\omega=\omega-\omega_{cav}=\Gamma=\tfrac{1}{2}\Gamma_F$, one halfwidth above the TE$_{011}$ resonance, $\vec{E}_0(\vec{r})$ and $\vec{H}_0(\vec{r})$ satisfy the boundary conditions of a perfectly conducting cavity and $\alpha(\Gamma)=1$, as explicitly treated in [6,10]. The azimuthally symmetric $\vec{E}_0(\vec{r})$ is 0 on the cavity walls and the boundary condition for the wall losses dictates that the phase of the electric field must be $\pi/4$, which is satisfied when $\alpha(\Gamma)=1$. To tune a weakly coupled cavity with a uniform surface resistance to resonance, we shrink the diameter and height by $\delta$. Then, $\alpha(0)=0$ and $\vec{E}_0(\vec{r})$ does not change since the frequency at which we drive the cavity is constant. However, at the new position of the walls, $\vec{E}_0(\vec{r})$ is no longer 0, but instead is equal to $\vec{f}(\vec{r})$.[*] In this way, the phase of the total field on the metallic walls continues to be $\pi/4$. At the feed, the imaginary part of $\vec{f}(\vec{r})$ also goes to 0, consistent with a smaller driving field feeding a cavity tuned to resonance. If the losses are not homogeneous, or if there are large couplings, again $\vec{E}_0(\vec{r})$ changes on the boundaries and here will more generally have a small perturbation to its form such that it reproduces the contributions to $\vec{f}(\vec{r})$ on the boundaries due to losses, but not the contributions to $\vec{f}(\vec{r})$ due to incident driving fields. If the cavity is further shrunk so that $\Delta\omega=-\Gamma$, then we have $\alpha(-\Gamma)=-1$ and, everywhere on the cavity walls except at the feeds, $\vec{E}_0(\vec{r})=2\vec{f}(\vec{r})$ so that the phase of the total field is still $\pi/4$ at the walls. Note that as the cavity is tuned, the phase of the driving field relative to the phase of the standing wave changes as $\tan(\Delta\phi)=\alpha(\Delta\omega)$ and the amplitude of the feed increases to preserve the same $\vec{E}_0(\vec{r})$. In general, $\alpha(\Delta\omega)=\Delta\omega/\Gamma$ and we can take $\vec{E}_0(\vec{r})$ and $\vec{H}_0(\vec{r})$ as the solutions for perfectly conducting walls [10]. Provided that $\Delta\omega/\omega$ is small, of order $\delta\,\omega/c$ where $\delta$ is the skin depth, the errors in the fields produce corrections of order $\delta^2$ and we neglect these. In Section 5 we show that a cavity detuning can introduce a DCP error when the cavity is fed by multiple feeds with different phases.

---
[*] Unless another mode is near resonance, $\vec{f}(\vec{r})$ changes negligibly for small detunings.

We solve for the fields of clock cavities using finite element methods because holes in the endcaps excite many modes [10]. To understand the solutions and the power dependence of DCP errors, it is helpful to analyze a cavity with no endcap holes, for which the solution for the fields is analytic [10]. The primary $TE_{011}$ standing wave is:

$$\vec{E}_0(\rho,z) = \frac{\mu_0 \omega}{2} \frac{k_1}{\gamma_1} J_1(\gamma_1 \rho) \cos(k_1 z) \hat{\phi}$$

$$\vec{H}_0(\rho,z) = \frac{k_1}{2} \left( \frac{k_1}{\gamma_1} J_1(\gamma_1 \rho) \sin(k_1 z), 0, J_0(\gamma_1 \rho) \cos(k_1 z) \right)$$

(2)

where, for the $TE_{mnp}$ mode, $k_p = p\frac{\pi}{d}$, $\gamma_p = \sqrt{k^2 - k_p^2}$, $p=1,2,3,\ldots$, $k=\omega/c$, R is the radius of the cavity, $J_m(x)$ is the Bessel Function of the first kind, and the z component of $\vec{H}_0(r,z)$ is normalized so that $\int_{-d/2}^{d/2} H_{0z}(0,z)dz = 1$ with the endcaps at $z=\pm d/2$. In clocks, a static magnetic field is applied in the z direction so that only the z component of **H(r)** couples the two clock states. The phase of $H_z(\mathbf{r})$ is given by $\tan(\Phi)=-g_z(\mathbf{r})/H_z(\mathbf{r})$ and therefore $g_z(\mathbf{r})$ and $H_z(\mathbf{r})$ give the DCP errors.

Analytic three dimensional solutions for $\vec{f}(\vec{r})$, and $\vec{g}(\vec{r})$ can be written as a Fourier expansion in $\phi$ and z [10]. The phase of $H_z(\mathbf{r})$ is, to lowest order:

$$\Phi(\vec{r}) = \Phi_0 - \frac{\delta}{d} k_1 z \tan(k_1 z) + \sum_{m,odd\ p} \Phi_{mp} \left(\frac{\rho}{r_a}\right)^{m+2\delta_{m,0}} \frac{\cos(k_p z)}{\cos(k_1 z)} \cos(m\phi)$$

(3)

The first term is an arbitrary phase offset. The second term describes an azimuthally symmetric (m=0), longitudinal phase variation of order $\delta/d \approx 30$ μrad with no radial variation. Below we see that this term produces a negligible frequency shift at optimal power and a large frequency shift at higher powers. The coefficients $\Phi_{mp}$ in the last term are given in Table 1 for a weakly coupled cavity of radius R=26 mm and a typical aperture radius of $r_a$= 5mm. The m=1, p=1 term is the largest, corresponding to a linear phase gradient at the center of the cavity. It is normally minimized by feeding the cavity equally on opposite sides, either by balancing independent feeds [1, 3, 7, 11, 12, 20, 21] or with an external ring structure [2,22]. Since the phase variations are proportional to $\rho^m \cos(m\phi)$, they are negligible for large m so this Fourier series converges rapidly [10]. The longitudinal variation of these is $\cos(pk_1 z)/\cos(k_1 z)$. The p=1 modes have no phase variation in z. It directly follows that there is no power dependence of their DCP errors because the atoms always see a constant phase during any cavity traversal. For p≥3, the $\Phi_{mp}$'s become small, especially for cavities with radii that are much larger than the 20mm m=0 TE waveguide cutoff. For m=0, the dominant phase variation in the third term is proportional to $\rho^2$, which has no longitudinal variation. Note that the lowest order m=0 phase variation treated in [14] is proportional to $\rho^2 z^2$ and is not present in (3). In real cavities with endcap holes, the phase variations of the fields are often very nearly the same as (3) at the center of the cavity, far from the endcap holes. The fast convergence of the Fourier expansion in ϕ allows the 3D fields to be efficiently calculated as a short series, m≤4, of two dimensional finite element problems in ρ and z [10].

| $\Phi_{mp}$ (µrad) | p=1 | 3 |
|---|---|---|
| **m=0** | 5.2 | 0.16 |
| **1** | 92 | 0.35 |
| **2** | 32 | 0.16 |
| **3** | -11 | 0.06 |
| **4** | -0.5 | 0.017 |

**Table 1.** Contributions to the phase near the center of the cavity in microradians in (3) for a cavity of radius R=26mm that is resonant at 9.2 GHz. The scale of the m=0 p=1 longitudinal phase variation is δ/d, 27 µrad for this cavity.

## 3. MODEL OF FREQUENCY SHIFTS DUE TO DISTRIBUTED CAVITY PHASE VARIATIONS

In this section we describe our calculation of the change of the atomic transition probability due to the spatial phase variations Φ(**r**) of the microwave field. It is both desirable and possible to have minimal variations of the phase along all atomic trajectories through the cavity [10]. However, while the cavities in current clocks have small phase variations at their centers, nearly all have large phase variations near the holes in the cavity endcaps.

We perturbatively treat the small effects due to the loss field $g_z(\mathbf{r})$ which gives a simple and linear picture of DCP errors. The change in transition probability due to small phase variations can be treated with the sensitivity function s(t) [23]. For an atom following a trajectory **r**(t) [10]:

$$\delta P = -\frac{1}{2}\int_{-\infty}^{\infty}\Phi\left[\vec{r}(t)\right]\frac{ds(t)}{dt}dt \qquad (4)$$

In fountain clocks, the atoms interact with the cavity field for a short time during the upward and downward cavity passages and we therefore neglect the transverse motion of the atoms during cavity traversal. During the first, upward cavity passage, the sensitivity function is s(t)= −sin[θ(ρ$_1$,t)] sin[θ(ρ$_2$)] where θ(ρ$_1$,t) is the tipping angle as a function of time during the first passage. On the second, downward cavity passage, θ(ρ$_2$) is the total tipping angle, for which the overall microwave phase is shifted by π/2 [23]. As in Ref. [10], θ(ρ,t)=ηbv$_z$π/2 $\int_{-\infty}^{t} H_{0,z}\left[\vec{r}(t')\right]dt'$ where b is an amplitude scale factor and η≈1 such that the average of θ(ρ$_{1,2}$)≈ π/2 for b=1, as discussed in detail at the end of this section. The sensitivity function is similar for the second cavity passage [23].

To analyze the spatial variations of the phase, we change variables in (4) from an integral over time to an integral over position in the cavity. Here we neglect gravity during the cavity passages since the change in velocity is typically a small fraction of the velocity $v_z$ under normal operating conditions.[†] For an atom that traverses the cavity at $\mathbf{r_1}$ on the upward passage and returns downward through the cavity at $\mathbf{r_2}$, we get:

$$\delta P_m(b,\vec{r}_1,\vec{r}_2) = \tfrac{1}{2}\sin\left[\theta(\rho_2)\right]\delta\Phi_m(b,\rho_1)\cos(m\phi_1) \\ - \tfrac{1}{2}\sin\left[\theta(\rho_1)\right]\delta\Phi_m(b,\rho_2)\cos(m\phi_2) \quad (5)$$

where $\delta P(b,\mathbf{r_1},\mathbf{r_2}) = \sum_m \delta P_m(b,\vec{r}_1,\vec{r}_2)$ and the effective phase for a cavity traversal [10] is $\delta\Phi_{\text{eff}}(b,\rho,\phi) \equiv \sum_m \delta\Phi_m(b,\rho)\cos(m\phi)$ with:

$$\delta\Phi_m(b,\rho) \equiv -b\eta\frac{\pi}{2}\int_{-\infty}^{\infty}\cos\left[\theta(\rho,z)\right]g_{mz}(\rho,z)dz, \quad (6)$$

$\theta(\rho,z) \equiv \eta b\pi/2 \int_{-\infty}^{z} H_{0,z}(\rho,z)dz$, $\theta(\rho) \equiv \theta(\rho,\infty)$, and $g_{mz}(\mathbf{r}) = g_{mz}(\rho,z)\cos(m\phi)$. The definition of $\delta\Phi_m(b,\rho)$ in (6) gives $\delta\Phi_{\text{eff}}(b,\rho,\phi) = \Phi(\mathbf{r})\sin[\theta(\rho)]$ if $\Phi(\mathbf{r})$ does not depend on z. Equations (5)-(6) therefore reproduce the physically obvious result that applying a uniform phase shift on both passages produces no change in transition probability. Also, if an atom experiences the same tipping angle and arbitrary phase as a function of position on the two cavity passages (e.g. a retrace of its path), the phase distribution again produces no change in transition probability. Finally, at moderately high power, $\delta\Phi_m(b,\rho)$ oscillates with an amplitude that's relatively constant as the power increases if $g_{mz}(\rho,z)$ does not oscillate wildly because the oscillatory behavior of $\cos[\theta(\rho,z)]$ with z makes the integral in (6) scale as 1/b. In the next section, the $m\geq 1$ modes of cavities without holes exhibit this behavior for b>10. However, the endcaps produce large and high frequency variations in $g_{mz}(\rho,z)$ and when, at some power, $\cos[\theta(\rho,z)]$ oscillates with $g_{mz}(\rho,z)$, the power dependence can be very large as we show in Section 4. We note that (6) gives the correct $\delta\Phi_m(b,\rho)$ to first order in $g_{mz}(\rho,z)$ even when $H_{0z}(\mathbf{r})$ is 0 and $\Phi$ is $\pi/2$. The only restriction is that the pulse area of $g_{mz}(\mathbf{r})$ is small.

We now average (5) over the atomic trajectories. For density and velocity distributions that are uncorrelated when the atoms are launched, we get:

$$\overline{\delta P_m}(b) = \frac{1}{N}\int^{r_a}\int_{|\vec{r}_2-\vec{v}(t_2-t_1)|<r_a} n_0(\vec{r}_0)\left(\frac{1}{\pi u}e^{-(\vec{v}-\vec{v}_0)^2/u^2}\right)\delta P_m(b,\vec{r}_1,\vec{r}_2)W_d(\vec{r}_d)d\vec{v}d\vec{r}_2$$

$$N = \int^{r_a}\int_{|\vec{r}_2-\vec{v}(t_2-t_1)|<r_a} n_0(\vec{r}_0)\left(\frac{1}{\pi u}e^{-(\vec{v}-\vec{v}_0)^2/u^2}\right)W_d(\vec{r}_d)d\vec{v}d\vec{r}_2 \quad (7)$$

Here $\mathbf{r_{1,2,d}} = \mathbf{r_0} + \mathbf{v}\, t_{1,2,d}$, $t_{1,2,d}$ is the position at the time of first or second cavity passage or the detection time, $\mathbf{r_0}$ is the initial launch position, u is the most-probable thermal velocity, $\mathbf{v_0}$ is the average transverse velocity, the integration over the velocity is constrained by $\rho_1 = |\mathbf{r_2}-\mathbf{v}(t_2-t_1)|<r_a$, N is the number of detected atoms, and $W_d(\mathbf{r_d})$ is the detection probability at position $\mathbf{r_d}$. We do

---

[†] This is not a good approximation for short interrogation times. It can easily be included, along with the vertical spatial and velocity distributions of the atoms.

**Illustrative Density Distributions and Generic Fountain Parameters**

|   | $r_{00}$ (mm) | $\rho_{off}$ (mm) | $\alpha_{tilt}$ (mrad) | detection curvature a−b |
|---|---|---|---|---|
| **I** | **0.5** | **2** | 0 | 0 |
| **II**$_{(0,1,2)}$ | **0.5** | 0 | (0,**1**,0) | (0,0,**50%**) |
| **III** | **3** | **2** | 0 | 0 |
| **IV**$_{(0,1,2)}$ | **3** | 0 | (0,**1**,0) | (0,0,**50%**) |
| $\delta P_{(0,1,2)\delta}$ | 0 @ t=$t_1$ | (0,**2**,**2**) | 0 | 0 |

$t_1$=0.13 s   $t_2$= 0.63 s   $t_{detect}$= 0.7 s   $\Delta t_a$=0.035 s   $r_a$= 5 mm   T=1 µK

**Table 2.** Parameters for several illustrative density distributions and typical fountain parameters. Here $r_{00}$ is the 1/e cloud radius at launch, **$r_{off}$** is the transverse cloud position at launch ($\phi_{off}$ =0), $\alpha_{tilt}$ is the angular tilt of the entire fountain with respect to gravity, and the detection curvature describes the non-uniformity of the atomic state detection. The cutoff waveguide sections of the cavity give apertures of $r_a$ at $t_{1,2} \pm \Delta t_a$. For I and III, the launch direction is not vertical such that the cloud is centered on the downward cavity passage at $t_2$ whereas, for II$_n$ and IV$_n$, the launch direction is along the fountain axis, which is tilted for II$_1$ and IV$_1$.

not treat the usually small effects of the vertical spatial and vertical velocity distributions. There may be additional apertures for the atoms in the fountain, for example at detection or especially at the bottom of the cavity's cutoff waveguides. Any additional cuts are easy to implement in a Monte Carlo integration, which is what we normally do. Since (2)-(5) are linear in $g_{mz}(\mathbf{r})$, our Fourier decomposition of $g_{mz}(\mathbf{r})$ leads directly to a Fourier series in $\phi$ for $\overline{\delta P_m}$, for which the change in transition probability (or frequency error) due to each azimuthal Fourier component $\cos(m\phi)$ of the DCP error is simply summed to give the total $\overline{\delta P}$.

If the atomic cloud is infinitely large and infinitely hot, the densities of atoms on both passages are uniform and uncorrelated, and the average of $\delta P_m(b,\mathbf{r_1},\mathbf{r_2})$ over all trajectories in (4) is zero. More pragmatically, the initial cloud sizes of most clocks range from less than 1 mm, for MOT collection, to several millimeters for molasses collection. We therefore concentrate on two initial distributions – initial Gaussian distributions with 1/e halfwidths of 0.5mm and 3mm, comparable to the cavity apertures, as described in Table 2. Note that for molasses collection in a 1,1,1 geometry [2], the clouds can be larger than the cavity apertures. For the downward passage, the 1/e diameter of the cloud is almost always larger than the aperture.

A simpler concept for (7) is to write it as an average over the density distributions of the upward and downward cavity passages for all atoms that are detected [13]. However, correlations between $\mathbf{r_1}$ and $\mathbf{r_2}$, e.g. between $\delta\Phi_m(b,\rho_1)$ and $\sin[\theta(\rho_2)]$ in (5), are lost in this way. Nonetheless, this simplified "uncorrelated" model offers helpful insight and is usually very accurate, even up to moderate powers. With a uniform detection probability, we can define:

$$\overline{\delta P_{uc}} = \frac{1}{N^2} \int^{r_a} \int^{r_a} n_1(\vec{r_1}) n_2(\vec{r_2}) \delta P(b, \vec{r_1}, \vec{r_2}) d\vec{r_1} d\vec{r_2} \qquad (8)$$

This allows the four dimensional average over $\delta P_m(b,\mathbf{r_1},\mathbf{r_2})$ to be separated into products of averages over $\delta\Phi_m(b,\rho)$ and $\sin[\theta(\rho)]$. Any detection non-uniformity contributes principally via the strong correlation between $\mathbf{r_d}$ and $\mathbf{r_2}$. Since the time delay between the downward cavity passage and detection is short, a convenient and good approximation is $W_d(\mathbf{r_d}=\mathbf{r_2})$.

We can see the basic properties of DCP errors from (8) for large clouds that have small density variations. An initially large cloud can be described by a quadratic density variation $n_1(\mathbf{r_1})=n+n\ \alpha_1[(\mathbf{r_1}-\mathbf{r_{off}})^2/r_a^2-\frac{1}{2}]$ at the upward cavity traversal and similarly a curvature $\alpha_2$ for $n_2(\mathbf{r_2})$. Here, for simplicity, we have taken the launch direction to be vertical whereas the launch direction is usually adjusted to maximize the number of detected atoms. Inserting these into (8) and neglecting small terms $\alpha_1\alpha_2$, we see that the contributions from the uniform density terms cancel and only the difference in density curvatures on the two passages produces a DCP error if the correlations between $\mathbf{r_1}$ and $\mathbf{r_2}$ can be neglected.

$$\overline{\delta P_{m,uc}} = \frac{1}{N^2}\int^{r_a}\int^{r_a} n^2\left(\alpha_1-\alpha_2\right)\left[\left(\frac{\vec{r_1}-\vec{r}_{off}}{r_a}\right)^2 - \tfrac{1}{2}\right]\delta P_m\left(b,\vec{r_1},\vec{r_2}\right)d\vec{r_1}d\vec{r_2}$$
$$= \frac{n}{2N}(\alpha_1-\alpha_2)\left(\overline{\sin\left[\theta(\rho_2)\right]}\left[\int^{r_a}\left(\frac{\vec{r_1}-\vec{r}_{off}}{r_a}\right)^2 \delta\Phi_m(b,\rho_1)\cos(m\phi_1)d\vec{r_1}\right]\right.\quad (9)$$
$$\left.-\overline{\delta\Phi_m(b,\rho_2)\cos(m\phi_2)}\int^{r_a}\left(\frac{\vec{r_1}-\vec{r}_{off}}{r_a}\right)^2\sin\left[\theta(\rho_1)\right]d\vec{r_1}\right)$$

Alternatively, this can be viewed as a density distribution of $n_1(\mathbf{r_1})=n_1+n_1(\alpha_1-\alpha_2)$ $[(\mathbf{r_1}-\mathbf{r_{off}})^2/r_a^2-\frac{1}{2}]$ for the upward passage and uniform for the downward passage, $n_2(\mathbf{r_2})= n_2$.

The second term in (5) and (9) averages to zero for all $m\geq1$ phase variations for a cloud that is centered on the downward passage. Only the first term of (5), $\frac{1}{2}\sin[\theta(\rho_2)]\delta\Phi_m(b,\rho_1)$, contributes to $\overline{\delta P}$. In this term, $\overline{\sin\left[\theta(\rho_2)\right]}$ is straightforward Rabi flopping, albeit with an amplitude that decreases at high power. If the atomic cloud is not centered, $n_1(\mathbf{r_1})$ will have a $\cos(\phi)$ component which, when multiplied by an $m=1$ $\delta\Phi_1(b,\rho_1)\cos(\phi_1)$, will give a DCP error. For small clouds, $m\geq2$ could contribute but, for initially large atomic samples, $m\geq2$ density variations should normally be small. We can now also see that non-uniformities in the detection, $W_d(\mathbf{r_d})$, can produce $m=2$ DCP errors. We consider a quadratic detection probability $W_d(\mathbf{r_d})= (1-ax_d^2/r_a^2)(1-by_d^2/r_a^2) \approx 1-(a+b)\rho_d^2/2r_a^2 - (a-b)\rho_d^2/2r_a^2\cos(2\phi)$ where a and b can describe the detection laser beam intensity variation and imaging non-uniformities. Thus, unless the laser and imaging curvatures are equal, $a=b$, an $m=2$ DCP error results. If the laser and imaging is not centered on the cloud of atoms, it can also produce an $m=1$ DCP error and $a+b$ gives an $m=0$ contribution.

For azimuthally symmetric phase variations ($m=0$), both terms in (5) and (9) will generally contribute. For these, we are free to define $\Phi_0$ in (3) such that the average of $\delta\Phi_0(b=1,\rho)$ over a uniform density distribution is 0. Finally we note that the correlations between $\mathbf{r_1}$ and $\mathbf{r_2}$ can introduce an interesting effect. Even if $n_2(\mathbf{r_2})$ is uniform, the second term in (5) can be non-zero when averaged over the density distribution because $\mathbf{r_1}$ and $\mathbf{r_2}$ are correlated and $\sin[\theta(\rho_1)]$ could have the same $m=1$ azimuthal dependence as $\delta\Phi_1(b,\rho_2)\cos(\phi_2)$.

Small clouds offer the advantage of reducing collisional frequency shifts for cesium clocks [3,24], at the expense of potentially increasing DCP errors. A reasonably accurate simplification here is a delta function for the upward passage, $n_1(\mathbf{r}_1) = n_1\, \delta(\mathbf{r}_1 - \mathbf{r}_{\text{off}})$ and a uniform density for the downward passage. For a delta function distribution, (8) simplifies to:

$$\overline{\delta P_{m,\delta}} = \tfrac{1}{2}\overline{\sin[\theta(\rho_2)]} \sum_m \delta\Phi_m(b, \rho_{\text{off}})\cos(m\phi_{\text{off}}) - \tfrac{1}{2}\sin[\theta(\rho_{\text{off}})]\overline{\delta\Phi_0(b, \rho_2)}. \tag{10}$$

A small cloud that passes upwardly through the cavity off-center, $\mathbf{r}_{\text{off}} \neq 0$, will have an m=1 DCP error and can also give m=2 and higher contributions. The first term in (5) therefore gives nearly all of the m≥1 DCP error. The second m=0 term in (10) will generally be nonzero except at b=1, where the above definition of $\Phi_0$ insures that it is 0.

The frequency error of the clock at optimal power and its uncertainty are ultimately most important. To get the frequency shift $\delta\nu$ from the average change in the transition probability, we divide (7) by the slope of the Ramsey fringes. The slope for any atom is:

$$\frac{\partial P(\rho_1, \rho_2)}{\partial \nu} = -\frac{\pi}{2\Delta\nu}\sin[\theta(\rho_1)]\sin[\theta(\rho_2)] \tag{11}$$

where $\Delta\nu$ is the Ramsey fringe width. Averaged over the ensemble, the slope of the Ramsey fringes is:

$$\overline{\partial P / \partial \nu} = \frac{1}{N}\int^{r_a} \int_{|\vec{r}_2 - \vec{v}(t_2 - t_1)| < r_a} n_0(\vec{r}_0)\left(\tfrac{1}{\pi u}e^{-(\vec{v}-\vec{v}_0)^2/u^2}\right)\frac{\partial P(\rho_1, \rho_2)}{\partial \nu}W_d(\vec{r}_d)d\vec{v}d\vec{r}_2. \tag{12}$$

This gives a frequency shift of $\delta\nu = -\overline{\delta P}/\overline{\partial P/\partial\nu}$, which is approximately $2\overline{\delta P}\Delta\nu/\pi$ at b=1. Throughout the rest of the paper, we prefer to discuss $\delta P_m$ and $\overline{\delta P}_m$ because they are never singular. The frequency shift is singular whenever $\overline{\partial P/\partial\nu}$ passes through zero as the power is increased, near b= 2,4, …, and we show that measurements of $\overline{\delta P}$ at these powers are particularly sensitive to m=0 DCP errors.

We are now able to precisely discuss our normalization of the field amplitude η. We define a π/2 pulse as the first maximum of the Ramsey fringe slope for uniform and uncorrelated density distributions on both cavity passages. Here, we can use the wave equation to generally show that $\nabla_{tr}^2 \theta(\rho) = -k^2\theta(\rho)$ for any atomic trajectory that goes through the cavity if $\mathbf{H}(\mathbf{r})$ is azimuthally symmetric and goes to 0 outside the cavity. Therefore, we get $\theta(\rho)= \theta(0)J_0(k\rho)$, which for cesium is 0.78 θ(0) at ρ=5mm, independent of the cavity geometry. Note that this is a larger radial curvature than that for a pulse of $H_{0,z}(\mathbf{r}) \propto J_0(\gamma_1\rho)$ delivered at the center of the cavity. Explicitly, η is a solution of:

$$\frac{\partial}{\partial \eta}\int_0^{r_a} \sin\left[\eta\frac{\pi}{2}J_0(k\rho)\right]\rho d\rho = 0 \tag{13}$$

For $r_a$=5(6)mm, (13) gives η=1.120(1.175). As a result, since the density distributions in clocks are not uniform, especially for the upward cavity passage, the maxima of the Ramsey fringe slopes will not occur at b=1,3,5, …., but at slightly smaller b's, such as b≈0.93, 2.8, 4.65,… For the upward passage, the cloud experiences an average tipping angle that is greater than π/2 at b=1, as defined by (13), because it is small and nearly centered.

## 4. DCP ERRORS OF CAVITIES WITH NO ENDCAP HOLES

It is instructive to use the analytic solution for a cylindrical cavity with no endcap holes as a starting point to understand DCP errors and their power dependence. We first consider the power dependence due to the spatial phase variations and ignore the spatial variation of the tipping angle θ. We then show that the spatial variation of the tipping angle produces a large power-dependent shift for azimuthally symmetric (m=0) phase variations. These general behaviors occur in cavities with endcap holes which we discuss in the rest of the paper. Here we use the example of R=26mm, near the average of the radii used in fountain clocks which range from 21.5 mm to 30 mm [3,4,7,21], and consider atomic trajectories within a typical cavity aperture of radius $r_a$=5mm, as in Table 1. Note that for cavity radii much greater than the $TE_{01}$ waveguide cutoff radius R=20 mm, large p modes are highly suppressed at the center of the cavity for a midplane feed. However, the figures in Sections 6-7 show that the endcap holes excite high p modes and these can produce significant power dependent DCP errors.

To show the basic features of the power dependence of the DCP errors, we first neglect the radial variation of the tipping angle. By taking $H_{0z}(\rho,z)= H_{0z}(0,z)$ and the lowest order phase variation from (3), we get:

$$\delta\Phi_{eff}(b,\rho,\phi) = b\frac{\pi}{2}\int_{-d/2}^{d/2}\cos[\theta(0,z)]H_{0z}(0,z)\Phi(\vec{r})dz$$

$$\delta\Phi_m(b,\rho) = \Phi_{mp}\left(\frac{\rho}{r_a}\right)^{m+2\delta_{m,0}}\delta\Phi_{long,p}(b) \qquad (14)$$

$$\delta\Phi_{long,p}(b) \equiv b\frac{\pi}{4}\int_{-\pi/2}^{\pi/2}\cos\left[b\frac{\pi}{4}(1+\sin u)\right]\cos(pu)du$$

The first two terms of (3) do not contribute here because they have no transverse variation. Taking a delta function density distribution at $\mathbf{r_{off}}$ on the first passage and a uniform density for the second gives:

$$\overline{\delta P}_{m,\delta} = \tfrac{1}{2}\sin(b\pi/2)\delta\Phi_m(b,\rho_{off})\cos(m\phi_{off}) = \sum_p \Phi_{mp}\left(\frac{\rho_{off}}{r_a}\right)^{m+2\delta_{m,0}}\cos(m\phi_{off})\delta P_{mp,\delta}(b)$$

$$\delta P_{mp,\delta}(b) = \tfrac{1}{2}\sin(b\pi/2)\delta\Phi_{long,p}(b) \qquad (15)$$

We plot $\delta P_{mp,\delta}(b)$ for p=1, 3 and 5 in Figure 1(a) to show the power dependence of the DCP errors.

In Figure 1(a). we see that frequency shifts of p=1 modes have no power dependence. Here $\delta P_{mp=1,\delta}(b)$ has the same dependence on b as the Ramsey fringe slope, $\partial P/\partial \nu$. For p=3 and 5, the frequency error at optimal power is small and $\delta P_{mp,\delta}(b)$ peaks at large b when p is large. For b larger than those shown in Fig 1a, $\delta P_{mp,\delta}(b)$ asymptotes to a sinusoid with a constant amplitude. In cavities with large radii, the endcap holes excite p≥3 modes more efficiently than the feed, so the excitations are larger than the $\Phi_{m,p=3}$ coefficients in Table 1 suggest. When the spatial variation of the tipping angle θ is included, the curves in Figure 1a change little – $\delta P_{mp,\delta}(b)$ (and $\partial P/\partial\nu$) decrease slowly as b increases. In addition, the curves shift to slightly lower b because the atoms get a slightly larger tipping angle on the upward cavity passage and then the average tipping angle on the downward passage.

The azimuthally-symmetric (m=0) longitudinal phase gradient produces a large power dependent DCP error. Here it is essential to include the spatial variation of the tipping angle θ(ρ). This error is particularly interesting because $\overline{\delta P_{0,\delta}}(b)$ in (10) is largest at b=4, 8, and 12 [solid line in Figure 1(b)], corresponding to two 2, 4, and 6π pulses on each cavity passage, where there is nominally no Ramsey fringe contrast. The scale of $\overline{\delta P_{0,\delta}}(b)$ is 70 ppm, corresponding to the change in transition probability at optimal power due to a frequency shift of order 5×10$^{-15}$. But here, since ∂P/∂ν goes to zero, the frequency shift can be arbitrarily large. For this shift, the first term in (10) [dashed line in Fig. 1(b)], is proportional to $\overline{\sin[\theta(\rho_2)]}$ which is zero for a uniform density distribution at b=4,8, and 12 on the second passage since the average tipping angle is a multiple of 2π. For the second term (dotted line), a delta function atomic density distribution at ρ$_{off}$=0 for the first cavity passage gives sin[θ(ρ$_{off}$)]>0 at b=4,8, and 12 because the tipping angle is largest at the center of the cavity.[‡] Then, as depicted in the inset of Figure 1(b), the average of δΦ$_0$(b,ρ) is large at b=4, 8, and 12 because cos[θ(ρ,z)] (dashed line) and g$_{0z}$(ρ,z)≈ δk$_1^2$ z sin(k$_1$z)/2d − Φ$_0$ H$_{0z}$(ρ,z) (solid line) are both even functions of z. For b=2 (dotted line) and 6 however, cos[θ(ρ,z)] is odd in z and therefore δΦ$_0$(b,ρ) and $\overline{\delta P_{0,\delta}}(b)$ are nearly 0.

A key point is that the m=0 frequency shift is very well suppressed at optimal power (b=1). The longitudinal phase variation produces a shift but the velocity reversal of the fountain provides an accurate cancellation, as long as the Bloch vector rotates through a small angle. Thus, only m≥1 DCP errors contribute significantly at optimal power. Therefore, unless an m≥1 error is exaggerated so that it is larger than the m=0 error at b=4, it is difficult for power dependent measurements with the current cavities to probe the DCP errors that are relevant at optimal power. For cavities with two independent feeds, the m=1 term can be exaggerated with differential measurements by feeding power from only one feed or the other [20]. For small cavity radii, p≥3 modes have a greater excitation, but still the power dependence of these, the ratio of δP(b≠1) to δP(b=1) for m≥1, are much less in Figure 1(a) than those for m=0 in Figure 1(b). Therefore the m=0 DCP errors due to the endcap losses generally dominate the m≥1 DCP errors at high power, as well as nearly every other clock error. It is likely that these m=0 DCP errors are often confused with errors due to microwave leakage [14]. For power dependence to be most useful as a probe of a variety of clock errors, including microwave leakage [4] and the lensing of the atomic wavepackets [25], the longitudinal phase variations due to the endcap losses should be dramatically reduced. In Section 7, we show cavity designs that minimize phase variations with additional feeds that are not in the cavity midplane.

---

[‡] Since there are no endcap holes, the curvature of θ(ρ) is smaller, γ$_1^2$, instead of k$^2$ as for cavities with holes.

## 5. AMPLITUDE AND PHASE BALANCE OF CAVITY FEEDS

In this section we discuss feed imbalances and the phase variations, especially for strong cavity couplings. The solutions in [10] give the fields for cavities in terms of the boundary conditions on the cavity walls, specifically treating feeds that were infinitely narrow in $\phi$ and z. We note that even when cavities are strongly coupled, Q≪30,000, the feeds are still essentially infinitely small in the context of the Fourier expansions if the width in $\phi$ is much less than $\pi/m$ for m≤4. Similarly, feed heights are usually much less than the cavity height – when they are comparable [22], the corrections are small and easily calculable. In this section we first discuss strong coupling, analyze amplitude imbalances, and then phase imbalances, which show a novel DCP error that depends on the cavity detuning.

When the couplings that feed power into the cavity are large enough to significantly reduce the cavity Q, they not only provide power to the cavity but most of the power that they supply also leaves the cavity through the same couplings. The net power fed into the cavity from all of the feeds is equal to the total power that is absorbed by the walls, which is independent of the strength of the couplings. Therefore, the cavity fields do not change as the cavity coupling becomes strong unless there are imbalances between the feeds that modify the power flow through the cavity, which we now discuss.

*Amplitude Imbalances of the Feeds:* As in [10], we calculate the fields from the boundary conditions. We first caution that real feeds can have multiple reflections from feeding cavities and couplings so the power flow at the clock cavity wall can be significantly different than the microwave power that is intended for that feed. We find it often helpful to picture the boundary condition for strong couplings as very lossy conductors that surround each feed. In Figure 2(a), a large amount of power is fed in, and also leaves, at each feed . In [10] the fields at the feeds and the losses on the cavity sidewalls are decomposed into Fourier components in $\phi$, and also in z for the analytic solutions. Since $\mathbf{E_0}$=0 on the cavity walls, the dominant power flow, $\mathbf{S(r)}=\mathbf{E(r)}\times\mathbf{H(r)}$, is from $f_\phi(\mathbf{r})\hat{\phi}\times\mathbf{H_0(r)}$ and not $\mathbf{E_0(r)}\times\mathbf{g(r)}$ [10]. In Figure 2(b), we sketch the boundary condition for $f_\phi(\mathbf{r})$ for a weakly coupled cavity with symmetric feeds at $\phi$=0 and $\pi$ (solid) and then four additional contributions that describe overcoupling. Overcoupling implies greater losses at $\phi$=0 and $\pi$ and we characterize these (dotted lines) by their fractional change to the cavity Q, $Q_0/Q_1$ and $Q_0/Q_2$, respectively, where $Q_0$ is for weak coupling and the loaded Q is $(Q_0^{-1}+Q_1^{-1}+Q_2^{-1})^{-1}$. We allow the amplitudes of the 2 feeds to be different, $\xi_1 Q_0/Q$ and $\xi_2 Q_0/Q$, times the amplitude for a single weakly-coupled feed. Thus, maintaining the same standing wave field requires that $\xi_1+\xi_2=1$. By decomposing this boundary condition for $f_\phi(\mathbf{r})$ into azimuthal Fourier components, we get **f(r)** and **g(r)** for strong coupling:

$$\begin{pmatrix} \vec{f}_m(\vec{r}) \\ \vec{g}_m(\vec{r}) \end{pmatrix} = \left[ \xi_1 \frac{Q_0}{Q} - \frac{Q_0}{Q_1} + (-1)^m \left( \xi_2 \frac{Q_0}{Q} - \frac{Q_0}{Q_2} \right) \right] \begin{pmatrix} \vec{f}_{m,w}(\vec{r}) \\ \vec{g}_{m,w}(\vec{r}) \end{pmatrix}, \qquad (16)$$

in terms of $\mathbf{f_{m,w}(r)}$ and $\mathbf{g_{m,w}(r)}$, the fields of a cavity with a single weakly-coupled feed [10]. The two feeds excite odd m modes with opposite phases and therefore, if $\xi_1=\xi_2$ and $Q_1=Q_2$, there is no excitation of the odd m modes. This can be seen by integrating the boundary condition in Figure 2(b) over cos(m$\phi$). In the following sections we use this same picture for more than 2 feeds.

Equation (16) shows that reducing the Q by overcoupling can increase the phase gradients in the cavity. For even m, $\mathbf{g_m(r)}$ is always $\mathbf{g_{m,w}(r)}$. For odd m, we now discuss the most interesting amplitude imbalances.

*Single overcoupled feed:* $\xi_1=1$, $\xi_2=0$, and $Q_2 \to \infty$ yields $\mathbf{g_m(r)} = \mathbf{g_{m,w}(r)}$ and similarly for $\mathbf{f(r)}$. Additional power is fed in and also leaves at $\phi=0$, giving the same phase gradients as for a weakly coupled single feed.

*Single feed, two general losses:* $\xi_1=1$, and $\xi_2=0$ yields $\mathbf{g_m(r)}=[1+2(Q_0/Q_2)]\mathbf{g_{m,w}(r)}$. If the two losses are equal, $Q_1=Q_2$, (16) gives $\mathbf{g_m(r)}=Q_0/Q\mathbf{g_{m,w}(r)}$. The power flow across the cavity is larger due to the loss on the opposite side for odd m, but there is no change for even m, as compared to weak coupling. If the feed at $\phi=\pi$ is used instead, the sign of the DCP error reverses. This configuration is used to easily detect m=1 DCP errors [20].

*Two unequal feeds, two equal losses:* $\xi_1=½(1+\varepsilon/2)$, $\xi_2=½(1-\varepsilon/2)$, and $Q_1=Q_2$ yields $\mathbf{g_m(r)}= \varepsilon/2(Q_0/Q)\mathbf{g_{m,w}(r)}$. Thus, a 10% feed imbalance gives a 5% larger DCP error for odd m modes, scaled by $Q_0/Q$.

*Two equal feeds, two unequal losses:* $\xi_1=\xi_2=½$, $Q_1=2(1+\varepsilon/2)Q_L$, and $Q_2=2(1-\varepsilon/2)Q_L$ yields $\mathbf{g_m(r)}= \varepsilon/2(Q_0/Q-1)\mathbf{g_{m,w}(r)}$. Thus, a 10% loss imbalance gives a 5% larger error scaled by $Q_0/Q$. This describes cavities that are externally balanced. The balancing matches the power delivered from the two feeds, even if the losses are asymmetric [20]. Therefore, the enhancement is proportional to $(Q_0/Q-1)$ as the wall losses do not contribute. If the coupling is weak, the odd m DCP errors are suppressed to the extent that the feeds are balanced [12,21].

*Two unequal feeds, two unequal losses:* Here the feeds have the same imbalance as the losses: $\xi_1=½(1+\varepsilon/2)$, $\xi_2=½(1-\varepsilon/2)$, $Q_1=2(1-\varepsilon/2)Q_L$, and $Q_2=2(1+\varepsilon/2)$, giving $\mathbf{g_m(r)} = (\varepsilon/2)\mathbf{g_{m,w}(r)}$. Thus, a 10% loss imbalance gives a 5% larger error that is not scaled by $Q_0/Q$. This case describes cavities that are fed by multiple cavity apertures by a single external feed, for example by a ring waveguide around the outside of the cavity [2,22]. If the field in the surrounding waveguide is uniform, a larger coupling implies that more power goes in, but also more loss. Therefore the scale of the DCP errors does not increase as the cavity is strongly coupled.

Summarizing, overcoupling can lead to larger phase gradients in clock cavities and therefore larger DCP errors. The main advantage of overcoupling is reducing the sensitivity to temperature tuning [22], which can be particularly important for rubidium clocks using many atoms [1,26]. Feeding from either side allows the m=1 DCP errors to be exaggerated so that it is straightforward to minimize the errors by adjusting the atomic trajectories to produce no shift as discussed below. Feeding with a ring waveguide structure largely compensates for coupling differences, leading to much less degradation as the coupling increases. Below we suggest alternative cavity designs that have small m=0 DCP errors at elevated powers and m≥1 DCP errors that can be precisely evaluated.

*Feed Phase Imbalances:* Many believe that a phase difference between the feeds produces a power flow from one feed to the other and a larger m=1 DCP error [12]. However, some work has shown that phase imbalances produce a negligible phase variation [9]. With (16), we can superpose solutions for feeding from either side as depicted in Figure 3. We show that the resulting fields do not lead to larger phase gradients when the cavity is tuned to resonance. But, when the cavity has a large detuning [27], the sensitivity to phase differences of the feeds is enhanced. This shows that a DCP error can depend on the cavity detuning and explains the previous differences.

We begin by considering a cavity tuned to resonance, $\alpha(\Delta\omega)=0$ in (1). Superposing the solutions for two identical and balanced feeds with a relative phase shift of $\varphi$, (16) becomes:

$$\begin{pmatrix} \vec{f}_m(\vec{r}) \\ \vec{g}_m(\vec{r}) \end{pmatrix}_{\text{odd m}} = \frac{1}{2\cos(\varphi/2)} \left[ e^{i\varphi/2} \frac{Q_0}{Q} - e^{-i\varphi/2} \frac{Q_0}{Q} \right] \begin{pmatrix} \vec{f}_{m,w}(\vec{r}) \\ \vec{g}_{m,w}(\vec{r}) \end{pmatrix}_{\text{odd m}}, \qquad (17)$$

which simplifies to $\mathbf{g_m(r)} = i \tan(\varphi/2) Q_0/Q\, \mathbf{g_{m,w}(r)}$ for odd m, where again $\mathbf{g_m(r)} = \mathbf{g_{m,w}(r)}$ for even m. Here the key is that $g_{mz}(\mathbf{r})$ is imaginary and therefore is a standing wave with the same phase as $H_0z(\mathbf{r})$ so no power flows from one feed to the other. There are no phase gradients for odd m, just as for balanced feeding with equal phases.

When the cavity is detuned, phase imbalances change the power supplied by each feed and produce phase gradients. Equations (1) and (16) give $[\alpha(\Delta\omega)+i]\, g_{mz}(\mathbf{r}) = [-1+i\, \alpha(\Delta\omega)]$ $\tan(\varphi/2) Q_0/Q\, g_{mz,w}(\mathbf{r})$ for odd m. The phase imbalance therefore gives larger odd m phase gradients proportional to $\text{Im}[Hz(\mathbf{r})] = \text{Im}[(\alpha(\Delta\omega)+i)\, g_{mz}(\mathbf{r})] = (\Delta\omega Q_0/\Gamma Q)\tan(\varphi/2)g_{mz,w}(\mathbf{r})$. Here, the sign of this phase gradient reverses as a function of detuning – as the cavity is detuned, $f_{m\phi}(\mathbf{r})$ of one feed is closer to the "correct" phase that most effectively excites the cavity and more power flows from this feed into the cavity than the other since its contribution to the real part of $\mathbf{E(r)}$ is larger. Note that for a given detuning, this phase gradient is independent of both the coupling strength and the wall losses since $\Gamma Q$ is a constant ($\omega/2$) as also is $Q_0\, g_{mz,w}(\mathbf{r})$.

## 6. DCP ERRORS OF REAL CAVITIES

The cavities used in fountain clocks have machining imperfections and holes in the endcaps, to allow atoms to enter and exit. The holes produce several effects. First, they give a smaller longitudinal phase variation near the axis of the cavity since the phase does not go to $\pi/4$ at the endcap, as it does in a cavity with no holes. Second, near the aperture of the endcap holes, "the corners," the microwave fields are very large, as shown in Figure 4 (a) [10]. $H_{0z}(\mathbf{r})$ can be larger at the corner (inset) than at the center of cavity and the phase of the field approaches $\pi/4$, giving large phase variations. Third, the sign of the standing wave $H_{0z}(\mathbf{r})$ reverses in the cutoff wave guide near the walls [Figure 4(a)]. Finally, the aperture is a sharp waveguide discontinuity that excites a large number of modes, and thus, finite element calculations are efficient at solving for the fields. Machining imperfections, especially for example at the endcap holes and potential scratches in the cutoff waveguide sections, can also produce worrisomely large phase variations. Next we generally prove that locally large phase variations produce negligible DCP errors and then, in the rest of this section, discuss finite element solutions for real cavities and their DCP errors.

*DCP errors due to nearly singular fields at sharp corners:* Microwave fields are nearly singular at sharp corners such as the aperture of the endcap holes. $\mathbf{H_0(r)}$ and $f_\phi(\mathbf{r})$ diverge as $\rho^{-2/3}$ at a 90 degree corner and $\mathbf{g(r)}$ as $\rho^{-5/3}$ for distances as short as the skin depth or the surface's radius of curvature [10]. The fields can be even larger if the corner is sharper, for example, at any sharp machining irregularities in the cutoff waveguide sections. In Figures 4(b)-(e), we see that $g_{mz}(\mathbf{r})$ oscillates near the corner, where the losses are large, and also at the feeds, where power is supplied. This suggests that large phase variations due to any locally large loss or feed could average to zero to produce a small net effect.

We consider the transverse gradients of the effective phase, $\delta\Phi_{eff}(b,\rho,\phi)$. Here, of particular interest are the waveguide cutoffs and apertures and, more generally, any region where the atoms are close to a conducting surface where $\mathbf{H_0(r)}$ is not zero. In (6), we can treat $\cos(\theta)$ as constant whenever b is small. Further, near walls of the cutoff waveguide sections, the total tipping angle from the cavity entrance to 1 mm inside the cavity body is of order 2% and therefore $\cos(\theta)$ is essentially constant as long as b≲15. Thus,

$$\overline{\nabla}_{tr}\delta\Phi_{eff}(b,\rho,\phi) \approx -b\eta\frac{\pi}{2}\cos\left[\theta(\rho,z)\right]\overline{\nabla}_{tr}\int_{z_1}^{z_2}\vec{g}(\vec{r})\cdot\hat{n}_C dz. \tag{18}$$

Here we choose a coordinate system where the atomic trajectory is along $\hat{z}$ but allow the cavity to have an arbitrary orientation. We also allow for the static magnetic C field, nominally along $\hat{z}$, to have an arbitrary direction, $\hat{n}_C = \vec{n}_{C,tr} + n_{C,z}\hat{z}$. Hence $g_\rho(\mathbf{r})$ and $g_\phi(\mathbf{r})$ can produce phase variations and also DCP errors. Using the wave equation, we expand the transverse gradient in (18) as:

$$\mu_0\omega\overline{\nabla}_{tr}\int_{z_1}^{z_2}\vec{g}(\vec{r})\cdot\hat{n}_C dz = k^2\int_{z_1}^{z_2}\hat{n}_C \times \vec{f}(\vec{r})dz - \left(\hat{z}\times\overline{\nabla}_{tr}\right)\int_{z_1}^{z_2}\Upsilon(\vec{r})dz$$
$$+ \left[\left(\vec{n}_{C,tr}\cdot\overline{\nabla}_{tr}\right)\left(\hat{z}\times\vec{f}(\vec{r})\right) + \mu_0\omega n_{C,z}\vec{g}_{tr}(\vec{r})\right]_{z_1}^{z_2}. \tag{19}$$
$$\Upsilon(\vec{r}) = \vec{n}_{C,tr}\cdot\overline{\nabla}_{tr}f_z(\vec{r})$$

From (19), we can see that only the average losses throughout the cavity can produce DCP errors if $\hat{n}_C = \hat{z}$. The first term on the right hand side is not singular and produces DCP errors of order $k^2$ as in Figure 1. It is proportional to the average of the components of $\mathbf{f(r)}$ that are transverse to the C field [e.g. $f_\phi(\mathbf{r})$]. Skipping to the last surface term, it is 0 if b is small since $(z_1, z_2)$ can be $(-\infty,\infty)$, outside of the cavity. For moderate b, if $z_1$ and $z_2$ are not near a sharp edge in the cavity, for example $z_1=-\infty$ and $z_2$ is 1 mm from the endcap and inside the cavity body, this term is small because the gradients will have a scale of 1mm and therefore do not change significantly as $\rho$ approaches $r_a$.

In the second term in (19), $\Upsilon(\mathbf{r})$ is doubly suppressed. First, the C-field is nominally aligned with $\hat{z}$ and, second, $f_z(\mathbf{r})$ is nominally zero for TE modes, including m=0, and also for m≥1 modes since they should have small excitations. Tilting the fountain will change this suppression by changing the alignment of the C field and cavity axis relative to the atomic trajectory, especially on the second cavity passage. We can further show that this term is negligible if the atomic density, most relevantly on the downward passage, does not vary rapidly in $\phi$ near $\rho=r_a$. From (18)-(19), $\Upsilon(\mathbf{r})$ produces a contribution to $\delta\Phi_{eff}(b,\rho_2,\phi)$ of $\Phi_0 \sin(\theta)$ − b$\eta\pi/(2\mu_0\omega\rho_2)\cos(\theta) \int_0^{\rho_2}\int_{z_1}^{z_2} d\Upsilon(\vec{r})/d\phi\, dz dr$ where $\Phi_0$ is trivial. Neglecting correlations, (5) and (8) give $\overline{\delta P_{uc}} = -\sin[\theta(\rho_1)] \int_0^{r_a}\int_0^{2\pi} n(\vec{r}_2)\delta\Phi_{eff}(b,\rho_2,\phi)d\phi\rho_2 d\rho_2$. Wherever $f_z(\mathbf{r})$ varies rapidly in $\phi$, if $n(\vec{r}_2)$ is nearly constant, it can be taken outside of this integral and then the integral over $\phi$ of $d\Upsilon(\mathbf{r})/d\phi$ gives no contribution to $\overline{\delta P_{uc}}$. Note that machining burrs could protrude radially from the endcap hole or waveguide cutoffs and these could produce a locally rapid variation of the atomic density, a large phase variation, and perhaps a non-zero $\overline{\delta P_{uc}}$. Putting the most restrictive apertures for the atomic trajectories in the middle of the cutoff waveguides, as we suggest in Section 7, avoids this possibility. For all other calculations in this paper, we take $\hat{n}_C = \hat{z}$ and therefore the 2$^{nd}$ term in (19) never contributes.

Thus, in any region where the tipping angle is small and the atomic density varies slowly, including near the cavity apertures and in the cut-off waveguides, nearly singular fields do not produce DCP errors that are larger than those due to the average losses of the cavity walls. Nonetheless, there is no benefit to having conductors close to the atoms where $\mathbf{H_0(r)}$ is large. In the next section, we show that we can minimize DCP errors at higher powers when the cutoff waveguides and apertures are not close to any atomic trajectories where $\mathbf{H_0(r)}$ is appreciable.

*Finite element method:* We use finite elements to first solve for the eigenmode $E_{0\phi}(\mathbf{r})$ using the boundary condition of a perfect conductor, $E_{0\phi}=0$, and get $\mathbf{H_0(r)}$ from the curl of $E_{0\phi}(\mathbf{r})$, as in [10]. To solve for $\vec{f}(\vec{r})$ and $\vec{g}(\vec{r})$, we use the Fourier superposition in $\phi$ as in Section 2. The waveguide discontinuities preserve the azimuthal symmetries but mix a large number of p modes for each m. This sum of all p modes is the finite element solution of a two-dimensional problem in $\rho$ and z. For m=0, as in Figure 2, the boundary condition for $f_{0\phi}(\mathbf{r})$ is proportional to $\mathbf{H_0(r)}$ plus $f_\phi(\mathbf{r})$ at the cavity feeds. The feed amplitude naturally emerges from the finite element calculation of $f_{0\phi}(\mathbf{r})$ and the curl of $f_{0\phi}(\mathbf{r})$ gives $\mathbf{g_0(r)}$ [10]. Near a perfect corner for the endcap holes, one that is sharp on the scale of the skin depth, the fields diverge up to the scale of the skin depth [10]. Here, considering the methods used to manufacture cavities, we take a radius of curvature for the corner of 20 μm, about 30 δ, and see that the fields vary with this spatial scale [Figure 4(b)] instead of δ. Therefore we do not have to solve for the fields in the copper walls and we simply use the conductivity, as we do for the rest of the cavity. The preceding proof shows that these, and any other fast spatial variations that produce large phase gradients, produce no additional DCP error for moderate b.

For m≥1 we cannot simply solve for $f_{m\phi}(\mathbf{r})$ because the solutions are not purely TE. If the conductivity of the cavity walls is isotropic, the boundary condition for $f_{m\phi}(\mathbf{r})$ is 0 everywhere except at the feed (Figure 2), so only the feeds excite m≥1 modes and isotropic wall losses do not [10]. Regarding the coupling of TE and TM modes, this occurs wherever the conductor normal has a component in both the $\hat{\rho}$ and $\hat{z}$ directions, for example, at any corner. The usual boundary condition for the parallel component of $\mathbf{f_m(r)}$ to be 0 couples $f_{m\rho}(\mathbf{r})$ and $f_{mz}(\mathbf{r})$, thereby coupling TE and TM modes. We solve for the m≥1 modes by simultaneously solving for all 3 components of $\mathbf{g_m(r)}$, using both nodal and edge elements [28-30]. We describe our method in the Appendix.

In this way the field can be calculated for a wide variety of cavity shapes for any given boundary conditions. The more difficult part is being sure of the boundary conditions. How is power actually fed into the cavity? Is there a spatial variation of the cavity's surface conductivity? Some limits can be set on these by measuring the cavity Q if the cavity is weakly coupled, but these, while helpful, are so far not very stringent. Future measurements of DCP errors in a variety of configurations, coupled with calculations of the fields and errors might shed light on the reasonable sizes for conductivity variations. We now discuss the DCP error for each Fourier component of $\phi$.

*m=0 phase variations:* DCP errors are shown in Figure 5a for the example cavity of R=26mm and apertures of $r_a$=5 mm. The behavior is similar to that for the analytic solution in Figure 1(b). Here, the smaller longitudinal phase variations reduce the magnitude of $\overline{\delta P_0}$ but the larger transverse curvature of the tipping angle compensates so that the frequency shifts are similar. There are 5 cases plotted in Figure 5a and, for each, the amplitudes that produce the maximum Ramsey fringe slopes (Figure 5(b)) are denoted by dots, nominally at b=1,3, 5, … . The solid line (I) shows a small cloud, 1 mm 1/e diameter that is reasonably approximated by the simple delta function/uniform down model (dashed-dotted). The frequency shifts are slightly smaller for a cloud that is offset by 2 mm from the cavity center ($II_0$, dashed) and even smaller for a large centered cloud, 6 mm 1/e diameter ($IV_0$, dotted). The correlations between the two passages are small, even at the highest powers for the large cloud ($IV_0$, dotted vs. $IV_{0,un}$, dashed-double dotted). The inset shows that the shifts are more than 200 times smaller at optimal power, less than $10^{-17}$.

We emphasize that these power dependent frequency shifts, which are of order $10^{-14}$, must be present in all clocks with mid-plane feeds. These DCP errors are due to the phase gradients from endcap losses and they are unchanged for weak or strong coupling or when multiple feeds around the cavity midplane are used [7]. Note that the templates for these are very different from those in Ref. [14] which analyzed a field with a different form. We emphasize that the transverse variation of the tipping angle $\theta(\rho)$ must be included to reproduce the DCP errors in Figure 5(a). To evaluate other systematic frequency errors at high power, b≥5, the longitudinal phase variations should be reduced or differential measurements with reasonable immunity must be made. In the next section, we show that these longitudinal phase variations can be eliminated by using additional feeds that are not at the cavity midplane.

*m=1 phase gradients:* The m=1 Fourier component represent a phase gradient that occurs when power flows across the cavity. A cloud offset or a tilt of the fountain breaks the reflection symmetry of the atomic density distribution and will generally lead to a DCP error for any m=1 phase variation. In Figure 5c we show the m=1 errors for a cavity with a single feed. An offset of the cloud produces an m=1 DCP error because the average phase for the cloud on the upward passage is essentially given by the phase at the offset position whereas the mean effective phase for the downward passage is 0. Therefore, until the cloud size is significantly larger than the cavity aperture, the m=1 DCP error is relatively unaffected by the cloud size. At b=1, $\overline{\delta P_1}$ is 13 ppm as roughly expected from ½$\Phi_{11}$×(2mm/5mm) from Table 1. If power is supplied to only one of 2 strongly coupled feeds and the Q is ¼ of the weakly coupled Q, the errors are 4 times larger than those in Figure 5c. When fed alternately from one feed or the other, the observed frequency difference is another factor of 2 larger than those in Figure 5c. Therefore, frequency differences can be of order $10^{-14}$[20].

Tilting the fountain [20] gives a transverse acceleration to the atoms in a frame that tilts with the fountain. An m=1 error occurs because most atoms that survive the aperture of the cutoff waveguides experience a transverse acceleration, and therefore a net displacement, between the upward and downward cavity passages. In Figure 6 we show the density distribution of all atoms that survive the aperture of the lower waveguide cavity cutoff for the upward and downward cavity passages. Here we illustrate a point source and a small tilt. For large tilts or large clouds, other apertures may truncate the density distribution. On the upward passage, the transverse acceleration has not yet significantly displaced the cloud center in Figure 6(a).

However, many atoms with $x_1<0$ will be cut by the lower cutoff waveguide and therefore the mean position of the atoms is shifted to positive $x_1$. For the downward cavity passage, gravity has accelerated the peak of the distribution in Figure 6(b) to negative $x_2$. This off-center density nearly fills the cavity aperture because the atoms will not spread much more before reaching the lower cutoff waveguide. Therefore, the mean effective phase on the upward passage is positive and negative for the downward, and they contribute nearly equally to the m=1 DCP error.

As suggested in Figure 1(a), the m=1 power dependence in Figure 5c is small. Both $\overline{\delta P_1}$ and the Ramsey fringe contrast decrease as b increases. However, many cavities have "$TM_{111}$ mode filters" which bring a TE mode closer to resonance. Its coupling with the nearby $TM_{111}$ mode can increase the DCP error in Figure 5d at optimal power. Due to the scattering by the endcap holes, it can produce a significant power dependence.

The large scale for the m=1 DCP errors at optimal power can be worrisome. For fountains with independent opposing feeds [1,3,7,11,12,20,21], it is straightforward to balance the feeds and to align the fountain to have no net combination of tilt and offset along the feed axis by alternately feeding from opposing sides and nulling the shift. If the shift is zeroed to less than $10^{-15}$ and then the two feeds are balanced to 1%, the m=1 DCP frequency uncertainty is less than $10^{-17}$. Without independent opposing cavity feeds, e.g. a ring waveguide feed [2,22], and perpendicular to the axis of independent feeds, it can be difficult to accurately assess the cloud position on the two passages. If the cavity is weakly coupled, the feeds are balanced to 1%, and the cloud offset is less than 2mm, Figure 5c shows that the m=1 DCP error is of order $10^{-17}$. If the cavity is strongly coupled, m=1 DCP errors can be correspondingly larger. How identically are feeds typically machined remains an unanswered question. In addition, it has not been shown that the wall resistance is homogeneous on the scale of 1% from one side of the cavity to the other. For ring waveguide feeds, the m=1 power dependence is unfortunately dwarfed by the m=0, so power dependence alone cannot be used to set m=1 DCP error limits. In the next section we show that not only can m=0 longitudinal phase variations be corrected, but also that the mode structure can be engineered so that m=1 DCP errors are small at optimal power and have a large power dependence.

*m=2 phase variations:* The m=2 phase gradients are potentially important for cavities with 2 feeds at $\phi=0$ and $\pi$. The gradients describe power going in at the feeds and being absorbed at $\pm\pi/2$ and are not enhanced by strong coupling. A large cloud is unlikely to have a significant m=2 density variation but a small cloud that is off-center, e.g. a delta function, will experience a non-zero m=2 effective phase on the upward passage, as shown in Figure 5(e) (gray dashed-double dotted). As discussed in Section 3, the atomic state detection can also produce an m=2 DCP error. The detection laser beams are almost always perpendicular to the imaging axis and, in many clocks, one of these axes is nearly aligned with the cavity feeds. Detection non-uniformities will preferentially detect atoms at $\phi=0$ and $\pi$ or $\phi=\pm\pi/2$, giving an m=2 DCP error. The m=2 detection related DCP errors can be suppressed by orienting the cavity so that the feed axis is at 45°, halfway between the detection laser propagation axis and the imaging axis. In Figure 5(e) we show calculations for a quadratic variation of the detection probability along the direction of the feeds with a−b=50%. While the m=2 DCP errors show modest power variations, they too are small in comparison to the m=0 power dependence. If a cavity has more than 2 feeds uniformly distributed in $\phi$ and the feeds are balanced to better than 1%, the m=2 DCP errors can easily be less than $10^{-17}$.

*m≥3 phase variations:* Phase variations for m≥3 can almost always be neglected. It is easy to design the cavity so that higher m modes are not resonant and then Table 1 shows that they produce small phase variations at the center of the cavity. The m=3 modes are nominally not excited for 2 or 4 mid-plane cavity feeds and, if only one feed is used, the m=1 phase gradients dominate. For m≥4, unless the cloud is small and has a large offset, these DCP errors are easily controlled to less than $10^{-17}$. The power dependences for m=3 and 4 generally resemble those for m=2.

## 7. MINIMIZING PHASE VARIATIONS IN CAVITIES

In this section we show several techniques to minimize phase gradients in cavities to arbitrary precision. We give an example that can give DCP errors below $10^{-17}$ for reasonably high powers, b<10. A basic principle is to locally feed the amount of power that is absorbed at each point in the cavity [10]. One novel concept is a cavity with a wall thickness of several skin depths that is enclosed inside another cavity with the correct mode shape to feed it appropriately. Aside from potential construction challenges, a greater difficulty is likely to be demonstrating that the cavity field has no phase variations. It is good but not sufficient to have a cavity with no phase variations – the cavity design should allow the phase variations to be evaluated. Here, our approach follows from realizing that since only a few modes produce significant DCP errors, a relatively small number of discrete feeds are required to minimize the errors. These solutions also allow the evaluation of the DCP errors. With a small number of controllable feeds, DCP errors can be measured to validate the models of the phase variations, null the cloud offset and fountain tilt, and thereby evaluate the DCP contribution to the uncertainty of the clock's frequency.

Our first goal is to minimize the azimuthally symmetric (m=0) DCP errors due to longitudinal phase gradients. As above, these produce a small frequency shift at optimal power and large shifts near b=4 and 8. Minimizing these is important for using power dependence to evaluate other frequency shifts, including the microwave leakage [4] and the microwave lensing of the atomic cloud [25], which has also been called a microwave photon recoil shift. If the DCP errors at high power are to be small, there can be no fast variations of $g_{0z}(\mathbf{r})$. This implies that the cavity walls should be far away from the atoms except where the standing wave field $H_0(\mathbf{r})$ is nearly zero. Using cutoff waveguides with a large diameter sufficiently attenuates $H_0(\mathbf{r})$ so that the usual cutoff waveguide can follow [10]. In Figure 7, we use cutoff waveguides with a 21mm [31.5mm] diameter in (a) [(b)] to attenuate $H_0(\mathbf{r})$. For both, the second cutoff waveguide diameter is 16.8mm, just below the m=1 cutoff, to slowly attenuate m=1 modes. This is followed by a 13 mm diameter cutoff section, which strongly attenuates all modes, and it includes a short 12 mm diameter aperture for the atoms in the middle of it. Therefore atoms are always more than 0.5 mm away from the cavity walls, except at these apertures where $H_0(\mathbf{r})$ is vanishingly small. Secondly, the feed positions and the losses in the cavity can be arranged to minimize the m=0 longitudinal phase variations. This implies cavity diameters closer to the m=0 TE cutoff, nominally less than 22 mm, so that p>1 modes can be easily excited [10]. For large cavity radii, small feeds will excite p=3 and higher modes but these are strongly attenuated as they propagate radially. Thirdly, the mode structure for m=1, and also m=2, can be tuned with minimal perturbation of the m=0 modes. This can minimize the DCP error at optimal power and give the m≥1 modes a moderate power dependence that allows an efficient evaluation of the DCP errors.

In our first example, we use the improved cavity from [10] which has feeds at two longitudinal planes. The idea in Figure 7(a) is to feed power closer to the endcaps instead of at the cavity midplane. Thus, the power flows on average radially inward instead of flowing longitudinally from the midplane to the endcaps. The distance of the feeds from the midplane is one free parameter and therefore we are able to only cancel the z tan($k_1$ z) phase variation in (3) that is responsible for the large m=0 DCP error at b=4, in Figure 8(a). By cancelling $\overline{\delta P_0}(b=4)$, the DCP errors at b=3 and b=5 are naturally small, a reduction by a factor of about 50 as compared to midplane feeds. Here, $\overline{\delta P_0}(b<6) \lesssim 1.5$ ppm [Figure 8(b) insets] corresponds to frequency shifts of order $10^{-16}$, and below $10^{-17}$ at optimal power. At higher power, the shifts are also smaller because the predominant phase variations have been canceled and also because the field has no fast variation since the corners are far from all atomic trajectories.

We minimize the m=1 and 2 DCP errors of this cavity by tuning the mode spectrum with the length of the 21mm diameter cutoff waveguide section and the top-wall outer cavity extension. By tuning higher modes (e.g. p=3), we can cancel the shift at optimal power for m=1 and 2 [10]. Even if this cavity is fed only at φ=0, the DCP errors at optimal power can be below $10^{-16}$. By feeding with two independent waveguide feeds at φ=0 and π [1,3,12,20,21], both of which supply equal power at z=±13.22mm, the m=1 DCP errors can easily be evaluated. The tilt of the fountain can be adjusted so that the average cloud position is the same on the two cavity traversals so the shift at b=3 is easily less than $10^{-15}$ [20]. Balancing the feeds at φ=0 and π to 1% reduces the shifts in Figure 8(b) by a factor of 200 so that they are below $10^{-17}$ at all b that have reasonable contrast. A better feeding scheme is to feed power to one waveguide at φ=±π/4 and the second at φ=±3π/4 [7,11]. This essentially eliminates m=2 phase variations while still allowing m=1 DCP errors to be evaluated along one tilt axis, for example, to confirm the suppression of the DCP error at optimal power of this cavity. Here, the m=1 power dependence is large enough that it could be fed at 4 φ's with a single ring waveguide structure [2,22]. The advantage of independent feeds is that the frequency shifts are much larger and easier to measure, which implies that there should be independent pairs of feeds along two orthogonal axes. With either configuration of more than 2 feeds, the m=2 adjustment extension could be eliminated. This cavity is quite long and the standing wave field $H_0(\mathbf{r})$ occupies only a fraction of the total length. The length could easily be shortened by allowing the diameter of 21mm m=0 waveguides to increase to as large as 36 mm beyond 1cm from the endcaps. As the cavity is drawn, the 12mm diameter cavity apertures cut the atoms at times greater then 35ms before and after they enter the cavity. Nonetheless, we use the parameters of table 2 so that the comparisons with Figure 5 are straightforward and because the differences are small and the cutoff sections could be shortened without significantly changing the DCP errors in Figure 8.

The longitudinal phase variations can be further reduced by adding more longitudinal feeds. Figure 7(b) shows a cavity with feeds at 4 heights. Here, the main body of the cavity can be fed by a ring waveguide at z=±7.9mm, φ=±π/4, ±3π/4 [2,22]. Regarding the power supplied to the m=0 cutoffs at z=±22.48, one possibility is to feed the bottom cutoffs at φ=±π/4 and independently at ±3π/4 so that the m=1 DCP errors can be cancelled for fountain tilts about the y axis. The top cutoffs could then be fed at φ=(π/4,3π/4) and independently at φ=(−π/4, −3π/4) to cancel tilts about the x axis. Alternatively, the top cutoff could be fed by a ring waveguide and the main body by two independent feeds to evaluate x tilts. Either way, there are a total of 16 cavity couplings that are driven by 5 independent cables. Figure 9(a) shows that the DCP errors at both b=4 and b=8 can be dramatically reduced, by more than a factor of 100 by feeding 12% of the power to the m=0 cutoffs. For even as high as 7 times optimal amplitude, $\overline{\delta P_0}$ is less than 0.15 ppm, which corresponds to the population changes at optimal power for a frequency shift of $10^{-17}$.

It has been argued that longitudinally distributed feeds as in [10] and Figure 7 should not be used because power can flow from one feed to the other, producing large Doppler shifts in the cavity [17]. We emphasize that longitudinally distributed feeds *are required* to cancel the longitudinal phase variations that exist in all of the cavities currently used in atomic clocks that have midplane feeds. In the language of power flow, as long as the feeds are reasonably balanced and the coupling is not too strong, the longitudinal phase variations in Figure 7(a), and especially 7(b), will be far better than those in a cavity with a midplane feed. As in [10], here it is also important to check that even p modes (2, 4, 6, …), corresponding to power flow from the top half to the bottom half of the cavity, are not nearly resonant. In Figure 9(b), we show the m=0 DCP errors due to even p modes. The reasoning used in the inset of Figure 1(b) for b=4 implies that the DCP error for even p modes is largest at b=2, 6, 10 …. If there are 8 feeds in the cavity body which have random longitudinal imbalances of 1% for any pair, the DCP errors in Figure 9(b) are suppressed by a factor of 400 and never exceed 0.1 ppm. The waveguide feeds are more sensitive at high power, essentially because they do not couple so well to the cavity body. If the cutoffs are fed by 4 cables balanced to within 1%, the errors in Figure 9(b) are suppressed by a factor of $2^{½}\times200$, yielding negligible m=0 DCP errors at optimal power and, near b=6, potentially as large as 0.25 ppm. For this cavity, the power distribution between the cutoffs and the body can be adjusted by balancing the cutoff feeds near b=5.3 and then adjusting the power split between the waveguides and the main body near b=3.8. This procedure should yield the small power dependence of Figure 9(a).

As for Figure 8, we minimize the m=1 errors in Fig. 9 (c-d) by adjusting the length of the 35.5 mm diameter cutoff waveguide section. In Figures 9(c-d) the m=1 DCP errors are shown for odd and even p modes for feeds only at ϕ=0. These errors can be greatly suppressed by using multiple feeds in ϕ and by aligning the fountain tilt to minimize m=1 DCP errors. At optimal power, the cutoff even p modes in Figure 9(d) give a large tilt sensitivity that can be used to align the fountain tilt. For example, one could feed the bottom cutoff at ϕ=±π/4, giving an m=1 excitation. Then an equal power is fed to the top cutoff, but here uniformly distributed in ϕ, and no power to the body. In this way, the cutoff power is not reduced to 12% of optimal so the excitation of the m=1 even and odd p DCP errors are 8 times larger than those in Figure 9(c)-(d). We also excite the m=0 odd p modes but, importantly, not the m=0 even p modes. Thus, the m=1 even p modes give a sensitivity of 40 ppm/mrad whereas the m=1 odd p gives 1 ppm/mrad and the m=0 odd p only 4 ppm/rad. This allows the fountain tilt to easily be aligned to 0.1 mrad. Similarly, by feeding just the top cutoff at ϕ=π/4 and 3π/4 and the bottom uniformly in ϕ, the fountain tilt in the orthogonal direction could also be aligned. Subsequently balancing the feeds to 1% will reduce the DCP errors by an additional factor of 200, giving DCP errors less than 0.1 ppm, even at high power.

*Optimization Methods:* Here we outline the optimization procedure used to obtain the solutions in Fig. 7 and discuss some techniques to minimize DCP errors. We first want to be clear that these cavities are merely good solutions, albeit very good solutions. While the DCP errors are so small that further improvements might never be useful, they are almost certainly not the ultimate solutions. It is difficult to prove that any solution is the global minimum given the large number of parameters and many different cavity configurations. More importantly, it is currently unclear what would even define the optimum. How uniform can the surface resistance of cavities assuredly be? How well can feeds be balanced? If the DCP errors for the cavities in Figure 7 are so small that they are unimportant, the design could minimize other systematic errors, such as line-pulling from 50 or 60 Hz spectral impurities.

For Figure 7(a), we first restricted ourselves to feeds at two longitudinal planes in the body of the cavity that are symmetric about the midplane. We do a least-squares minimization of $\overline{\delta P_0}(b)$. Since we expect that we can only null the m=0 shift at b=4 and since the shift is expected to be small at b=6, we aspire to minimize $\overline{\delta P_0}(b)$ up to b=5 and therefore apply a weight to the chi square of $\overline{\delta P_0}(b)$ that linearly decreases to 0 at b=6. For the chi square minimization, we calculate $\overline{\delta P_m}(b)$ with m≤2 for the simplified model of delta function up and uniform down, with the cloud centered at r=0 and 3mm on the upward passage. The least-squares minimization varied 3 parameters, the cavity radius, the cutoff waveguide radius, and the feed position. We performed these minimizations for a variety of the radial gaps of the m=2 top-wall extension. We found a weak dependence up to 5mm and use this to give a large coupling. A small coupling requires the extension to be precisely tuned near an m=2 resonance. For each set of parameters, the cavity height is first changed to tune the cavity to resonance. Second, $\overline{\delta P_2}(b=1)$ (m=2, optimal power) is nulled by varying the height of the top-wall extension. Finally, the length of the first cutoff section is varied to null the $\overline{\delta P_1}(b=1)$ (m=1, optimal power). The chi square is slightly lower than that for Figure 7(a) for cavity radii as small as 20.85 mm, but those cavities are even longer. The 21.28 mm radius of the improved cavity design from [10] is a reasonable choice. The chi square is also slightly smaller if the first cutoff waveguide sections have a diameter less than 21 mm, but then this section would get very long since it would be near the m=1 cutoff diameter of 19.1 mm.

If we allow 4 longitudinal feed planes, we expect to minimize the large m=0 shifts at b=4 and 8 and we therefore used chi square weights that decrease linearly to 0 at b=10, instead of b=6 as above for 2 feed planes. To cancel the m=1 shift at optimal power from the main body feeds, the length of the first cutoff waveguide section can be tuned. Then, for many cavity geometries, a height of the waveguide feeds can be found for which they give no m=1 shift at optimal power. This height depends on the length of the first cutoff waveguide section, and also on its diameter as well as the diameter of the main cavity body. The minimum solution we found in this way had m=0 errors that are roughly 3 times larger than those in Figure 9(b). While it seems that a cavity shape such as this one does exist that allows the m=0 DCP errors to be small and the m=1 errors to be 0 at optimal power, this condition is too restrictive. It is only important that the m=1 error at optimal power be negligible (in addition to all shifts not being too large at all powers). We therefore add $\overline{\delta P_1}(b=1)$ due to both body and cutoff feeds to our chi square and adjust its weight so that its contributions are comparable to that for $\overline{\delta P_0}(b<10)$, giving Fig. 7 (b). Further, by adding and tuning TM mode filters at the cavity endcaps, as well as adjusting the length of the 31.5mm diameter waveguide sections, the m=1 DCP errors at optimal power from both the body and waveguide the feeds can be cancelled for this cavity. However, this solution gives a larger

m=1 power dependence than in Figs. 9 (c-d) – here, some compromise between reducing the m=1 shift at optimal power and the size of the shifts at high power may be preferable.

To find regions of local minima, we performed a Monte Carlo sampling of the large parameter space. We then used gradient searches to find the local minima. Many of these gradient searches showed that the optimal position for the waveguide feed is as close to the endcap as possible. This is difficult to construct so we imposed a minimum height of 3 or 6 mm above the endcap, and both gave reasonable solutions. For these searches, we used chi square weights that linearly decreased to zero at b=14. Since the shifts for b>10 were generally much larger, we subsequently focused only on the more useful region of b<10. There are a wide variety of solutions so factors other than DCP errors may be far more important. Here we note that we have restricted our searches to cavities that are symmetric about the midplane. This is an unnecessary restriction, especially since the sensitivity function is not symmetric about the midplane, including at b=1. We have also only considered feeds on the side walls and not multiple feeds on the top wall, which seems more difficult to construct. More experience with cavities and evaluating DCP errors is required to identify any future limitations.

## 8. CONCLUSION

The DCP errors due to phase variations in $TE_{011}$ cylindrical cavities can be naturally decomposed into Fourier components in $\phi$. The most important Fourier components are m≤2 and each produce DCP errors with unique characteristics that impact how the DCP uncertainty is evaluated, as well as its size. The m=0 DCP error is primarily due to longitudinal phase variations with no transverse dependence. They produce small errors at optimal power and large shifts at b=4 and 8, corresponding to two $2\pi$ and two $4\pi$ pulses. If the atomic cloud is small when launched, this implies that the frequency errors will be large for $5\pi/2$ and $9\pi/2$ pulses and not as large for $3\pi/2$ and $7\pi/2$ pulses. This m=0 shift could easily be misinterpreted as an other frequency shift, such as those due to microwave leakage. The m=1 DCP errors have a large scale size at optimal power and therefore, even for cavities with multiple feeds, uncertainty evaluations need to evaluate it. The m=1 DCP error is sensitive to the cloud position and therefore can be assessed with tilts along orthogonal axes. It nominally has a small power dependence but, if the cavity has "$TM_{111}$ mode filters," these often produce a large coupling to TE mode which can make the DCP error at optimal power larger and lead to a significant power dependence. Strong coupling naturally increases the m=1 DCP error, especially for independent feeds, and phase imbalances will produce a DCP error if the cavity is detuned. The m=2 DCP error is smaller, and should be negligible in cavities with 4 or more feeds. Here, an off-center, small, initial cloud produces an error as well as detection inhomogeneities. For m≥3, the DCP errors should be negligible. We have proven that sharp structures, while they produce locally large phase variations, do not produce anomalous DCP errors provided that the atomic density is slowly varying near the perturbation. It is important to experimentally verify the DCP errors that are calculated with finite element models. Verification will allow the models to be used to convincingly assess and probably reduce DCP uncertainties in the accuracy evaluations of current fountain clocks. Further, new cavity designs can have negligible phase variations which may allow smaller DCP uncertainties, especially for small clouds used to cancel the collisional frequency shift [24]. These cavity designs should also enable the reduction of other systematic uncertainties by enabling their evaluation at elevated microwave power.

## 9. ACKNOWLEDGEMENTS

We gratefully acknowledge stimulating discussions with S. Bize, J. Guéna, and R. Schröder and financial support from the National Science Foundation, NASA Microgravity program, the Office of Naval Research, and Penn State.

## 10. APPENDIX A: FINITE ELEMENT CALCULATIONS

We use finite element methods to solve the wave equation. For m≥1, we need to solve the wave equation for **g(r)** while simultaneously satisfying that the divergence of **g(r)** is zero. In contrast to m=0 TE modes, all 3 components of **g(r)** will generally be non-zero. For the m=0 calculations, we use nodal elements, where the solutions are determined at the nodes in the domain, for example, at the vertices of a triangular mesh. When solving for **g$_{m≥1}$(r)**, nodal elements give solutions, spurious modes, which do not have zero divergence. By using a combination of nodal and edge elements, the spurious modes are eliminated because the edge elements, by their construction, give solutions with zero divergence and also allow **g$_{m≥1}$(r)** to satisfy the boundary conditions [29,30]. We avoid singularities in the curl and divergence of the fields, and also impose the correct asymptotic form as ρ→0, by choosing:

$$g_{m\rho}(\vec{r}) = \left[\rho^m G_{m\rho}(\rho,z) - \rho^{m-1} G_{m\varphi}(\rho,z)\right]\cos(m\phi)$$
$$g_{m\varphi}(\vec{r}) = \rho^{m-1} G_{m\varphi}(\rho,z)\sin(m\phi) \qquad . \qquad (A1)$$
$$g_{mz}(\vec{r}) = \rho^m G_{mz}(\rho,z)\cos(m\phi)$$

Here, $G_{0\phi}(\mathbf{r})=0$ if m=0. We solve for $G_{m\rho}(\rho,z)$ and $G_{mz}(\rho,z)$ using edge elements and for $G_{m\phi}(\rho,z)$ with nodal elements, normally 3$^{rd}$ order elements for all. The edge elements force $g_{m\rho}(\mathbf{r})$ and $g_{mz}(\mathbf{r})$ to be continuous along the edge but may be discontinuous perpendicular to the edges, for example if the corner of the endcap holes is perfectly sharp. Everywhere, including near a sharp corner, $g_{m\phi}(\mathbf{r})$ is continuous in all directions, as given by nodal elements.

We cast the problem in the weak form by taking a volume integral over $\vec{g}_m^t(\vec{r}) \cdot \left(\vec{\nabla} \times \left[\vec{\nabla} \times \vec{g}_m(\vec{r})\right] - k^2 \vec{g}_m(\vec{r})\right)$. Here $\vec{g}_m^t(\vec{r})$ is a test function of the same form as (A1), but with {G$^t_{m\rho}$(ρ,z) G$^t_{m\phi}$(ρ,z), G$^t_{mz}$(ρ,z)}. We expand the curls to get a volume integral over $\left[\vec{\nabla} \times \vec{g}_m^t(\vec{r})\right] \cdot \left[\vec{\nabla} \times \vec{g}_m(\vec{r})\right] - k^2 \vec{g}_m^t(\vec{r}) \cdot \vec{g}_m(\vec{r})$ and a surface integral over $\varepsilon_0 \omega \vec{g}_m^t(\vec{r}) \cdot \left[\hat{n} \times \vec{f}_m(\vec{r})\right]$, which gives the boundary condition. If power is only supplied on the sidewalls, the surface term is only non-zero for $G_{mz}(\rho,z)$. Integrating the volume and surface over φ gives the weak form in r and z. Additionally, we constrain $G_{m\rho}(\rho,0)=G_{m\phi}(\rho,0)=0$ at the cavity mid-plane for odd p modes and $G_{mz}(\rho,0)=0$ for even p.

**Figure 1.** (color online) (a) Power dependence of the transition probability due to p=1, 3, and 5 modes which have phase variations $\cos(pk_1z)/\cos(k_1z)$. The lowest order modes (p=1) give a DCP error that is proportional to the Ramsey fringe contrast and therefore the frequency shift is independent of the tipping angle, $b\pi/2$. For p=3 and 5(dashed, dotted), the phase has rapid longitudinal spatial variations which produce small errors at optimal power (b=1) and larger errors at high power. (b) Power dependence of the transition probability $\overline{\delta P_{0,\delta}}(b)$ (solid black line) due to the azimuthally symmetric longitudinal phase variations (m=0, p=1) in a cavity with radius R=26mm and a small cloud. The shift is largest at b=4, 8, 12, corresponding to two $2\pi$, $4\pi$, $6\pi$ pulses during the cavity traversals. The dashed blue (dotted violet) lines show the contributions to $\overline{\delta P_{0,\delta}}(b)$ from the 1$^{st}$ (2$^{nd}$) term in (10) with $\rho_{off}$=0. The inset shows $g_{0z}(z)$ (dashed) multiplied by 100 and $\cos[\theta(z)]$ for b=2 (dotted) and b=4 (dashed). At b=2, $\cos[\theta(z)]g_{0z}(z)$ is an odd function of z and so its integral, the effective phase $\delta\Phi_0(b=2,\rho)$, is zero. At b=4, $\cos[\theta(z)]$ $g_{0z}(z)$ is an even function and therefore $\delta\Phi_0(b=4,\rho)$ and $\overline{\delta P_{0,\delta}}(b=4)$ are large.

**Figure 2.** (color online) (a) Model of feed amplitude imbalances. The cavity has wall losses, $Q_0$, and coupling losses at the feeds $Q_1$ and $Q_2$. Power is fed at 2 or more positions with amplitudes proportional to $\xi_1Q_0/Q$ and $\xi_2Q_0/Q$, with $\xi_1+\xi_2=1$. (b) Boundary condition for $f_\phi(\mathbf{r})$. The black line illustrates $f_\phi(\mathbf{r})$ for a weakly coupled cavity with 2 feeds. Larger feed couplings require more power to be supplied and asymmetries can lead to larger phase gradients.

**Figure 3.** Model of feed phase imbalances. The solution is a superposition of two solutions for a cavity fed by a single feed with different phases.

**Figure 4.** (color online) Finite element calculation of fields for a typical cavity, R=26mm and $r_a$=5mm. (a)-(e) show $H_{0z}(\mathbf{r})$, $g_{0z}(\mathbf{r})$, $g_{1z}(\mathbf{r})$, and $g_{2z}(\mathbf{r})$. The insets show a 50μm × 50μm region near the aperture of the endcap hole where the fields are nearly singular. The scale for $H_{0z}(\mathbf{r})$ is about $10^4$ larger than for $g_{mz}(\mathbf{r})$. In (d) the cavity has an m=1 mode filter that is commonly used. It significantly changes the field at corners (inset), which couples the m=1 modes. All external corners have radii of 20μm.

**Figure 5.** (color online) (a) Azimuthally symmetric DCP errors as a function of microwave amplitude for the cavity in Figure 4 and the illustrative density distributions in Table 2. The shift is below $10^{-17}$ at optimal power (inset) and large at b=4, 8, and 12, as in Figure 1(b). The simple delta function distribution (gray, dashed-double dotted) gives the scale of the errors and the uncorrelated average (7) is a good approximation, even for a relatively large cloud (IV$_{0,un}$ dash-dot-dot vs. IV$_0$ (dotted). (b) Ramsey fringe slope as a function of microwave amplitude b. The contrast decreases at high power due to the radial variation of the tipping angle and has a weak dependence on the density distribution. (c) m=1 DCP errors for a cavity with a single feed. The errors have a small power dependence and, if the cavity is fed by 2 or more feeds that are balanced to 1%, the errors are 100 times smaller. (d) m=1 DCP errors for a cavity as in (c) that has 7.5 mm long TM mode filters as in Figure 4(d). The mode filters increase the DCP errors at optimal power and produces a significant power dependence. (e) m=2 DCP errors for a cavity with 1 or 2 opposed feeds. A 2mm offset of a small cloud on the upward passage (solid black) gives a shift of order $2\times10^{-16}$ at b=1 and is slightly smaller for larger clouds (green, long dashed).

Detection inhomogeneities of 50% produce errors of order $10^{-16}$ (aqua dashed and pink dashed-dotted).

**Figure 6.** (color online) Density distributions ($\Pi_1$) of atoms that survive all apertures in a tilted fountain. On the upward passage (a), the peak of the distribution has not moved significantly but more atoms with positive $x_1$ will pass through the apertures. On the downward passage (b), the tilt has accelerated the peak of the cloud to negative $x_2$. The average difference is proportional to the tilt and couples with m=1 phase variations to produce a DCP error.

**Figure 7.** (color online) Cavities with minimal DCP errors. Power is fed at 2 [4] horizontal planes in (a) [(b)] to minimize the azimuthally symmetric (m=0) longitudinal phase variations. The contour plots of the x-z plane show the standing wave fields $H_{0z}(\mathbf{r})$ and $g_{mz}(\mathbf{r})$, which are symmetric in z. The contour steps are 10 m$^{-1}$ for $H_{0z}(\mathbf{r})$ and (0.2, 4, 2)$\times 10^{-3}$ m$^{-3}$ for quadrants 2-4 respectively in (a) and (0.04, 1, 2)$\times 10^{-3}$ m$^{-3}$ for (b). For (a), the large cutoff waveguide section and the endcap mode filter are tuned to cancel the DCP errors of $g_{1z}(\mathbf{r})$ and $g_{2z}(\mathbf{r})$ at optimal power. All external corners have radii of 0.1mm.

**Figure 8.** (color online) DCP errors for the improved cavity geometry in Figure 7(a). (a) Azimuthally symmetric DCP error (m=0) as in Figure 5(a). The error is less than $10^{-17}$ at optimal power (top inset) and, for b<6 (bottom inset), remains below 1.5 ppm. (b) m=1 DCP errors for a cavity fed at z=±13.22mm and ϕ=0. With feeds distributed at 2 or more ϕ's and balanced to 1%, the m=1 DCP error is below $10^{-17}$ for b<13. (c) m=2 DCP errors for a cavity with a feeds at ϕ=0 or ϕ=0 and π. The top-wall outer cavity extension is tuned to reduce the shift at optimal power (inset). If feeds at ϕ=±π/4, or evenly distributed at more than 2 ϕ's, are balanced to within 1%, the m=2 DCP errors are suppressed by at least a factor 100, less than $10^{-17}$.

**Figure 9.** (color online) DCP errors for the improved cavity geometry in Figure 7(b). (a) Azimuthally symmetric DCP error (m=0) as in Figure 5. The error is less than $10^{-17}$ at optimal power (inset) and generally remains below 0.15 ppm for b<10. Here the cutoff waveguide feeds supply 12% of the power. The DCP errors for body feeds ($I^b$, $II_0^b$, $IV_0^b$) with no power driving the cutoff feeds are ≈100 times larger. (b) m=0 DCP errors for even p modes that give longitudinal power imbalances. If power is fed only at z=+(−)7.9mm and +(−)22.48mm, the DCP error is the curve in (a) plus(minus) the 2 corresponding curves in (b). Power flowing from a top feed to the bottom feed sets up a phase variation that gives the largest DCP errors at b=2, 6, 10, … If feeds are balanced to 1%, these DCP errors are suppressed so the error is negligible at optimal power and remain small at high power. (c) & (d) DCP errors for m=1 for the body ($I^b$, $II_1^b$, $IV_1^b$) and cutoff feeds ($I^c$, $II_1^c$, $IV_1^c$) at ϕ=0 for (c) odd p and (d) even p. The cutoff m=1 even p modes can be used to align the tilt of the fountain. Balancing the feeds to 1% and aligning the fountain gives DCP errors less than 0.1 ppm at high power.

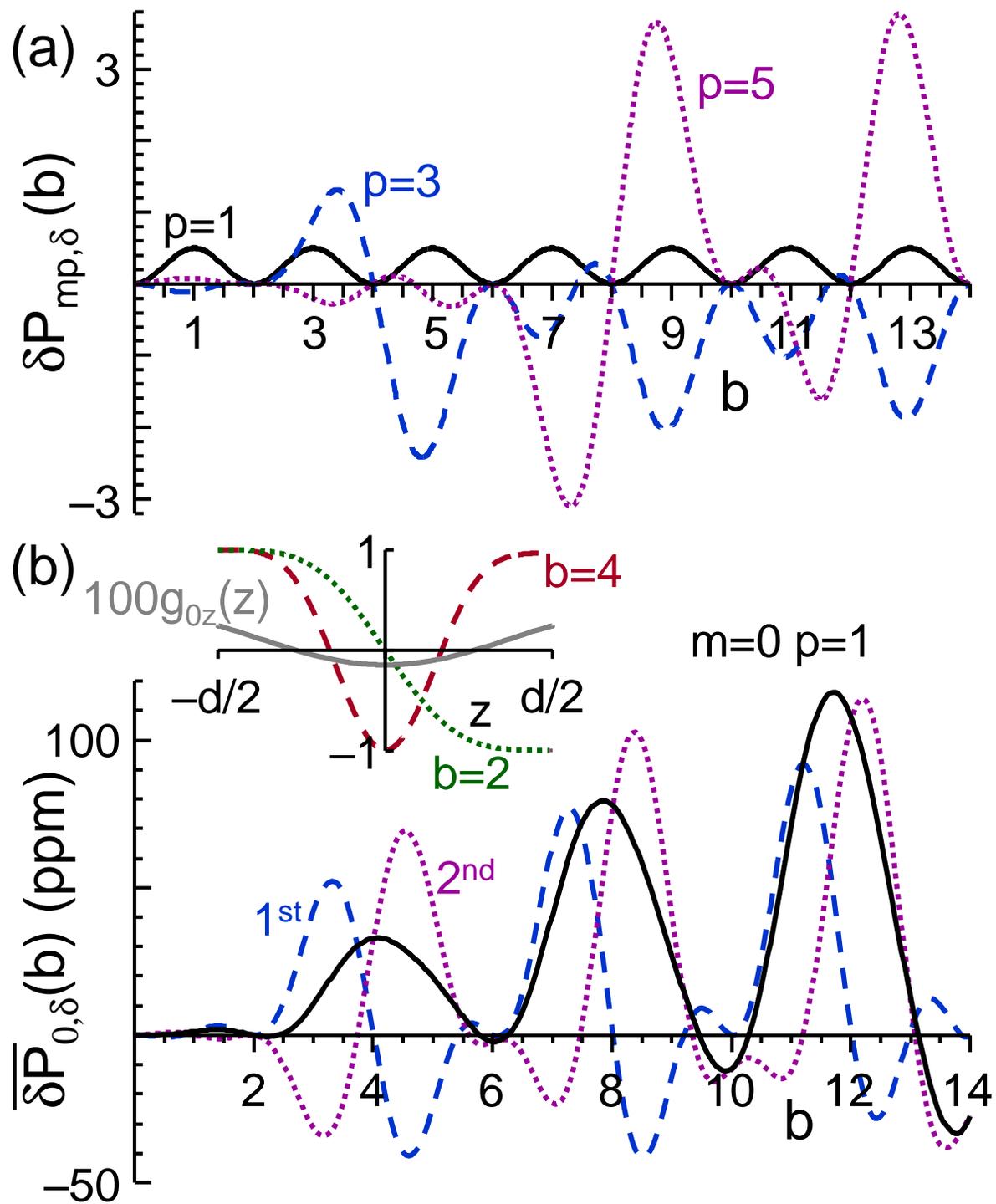

Li Fig. 1

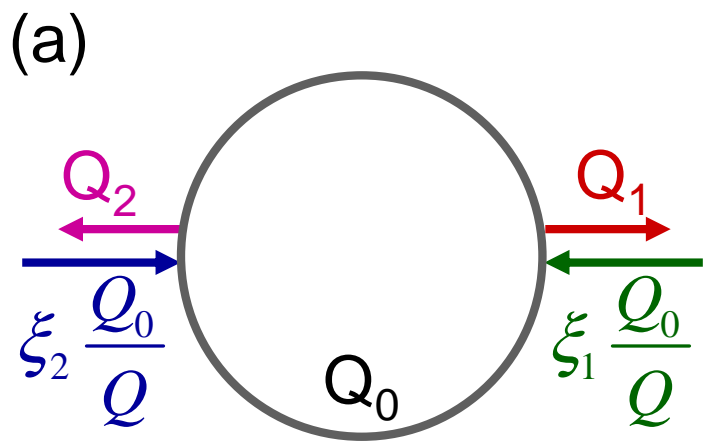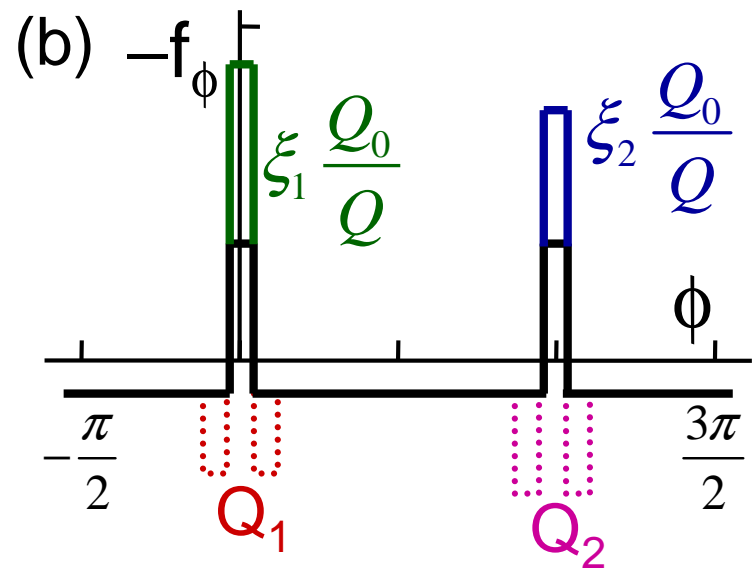

Li Fig. 2

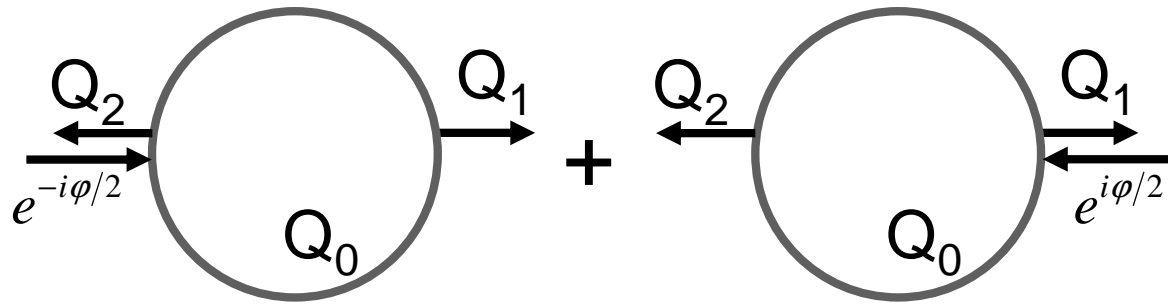

Li Fig. 3

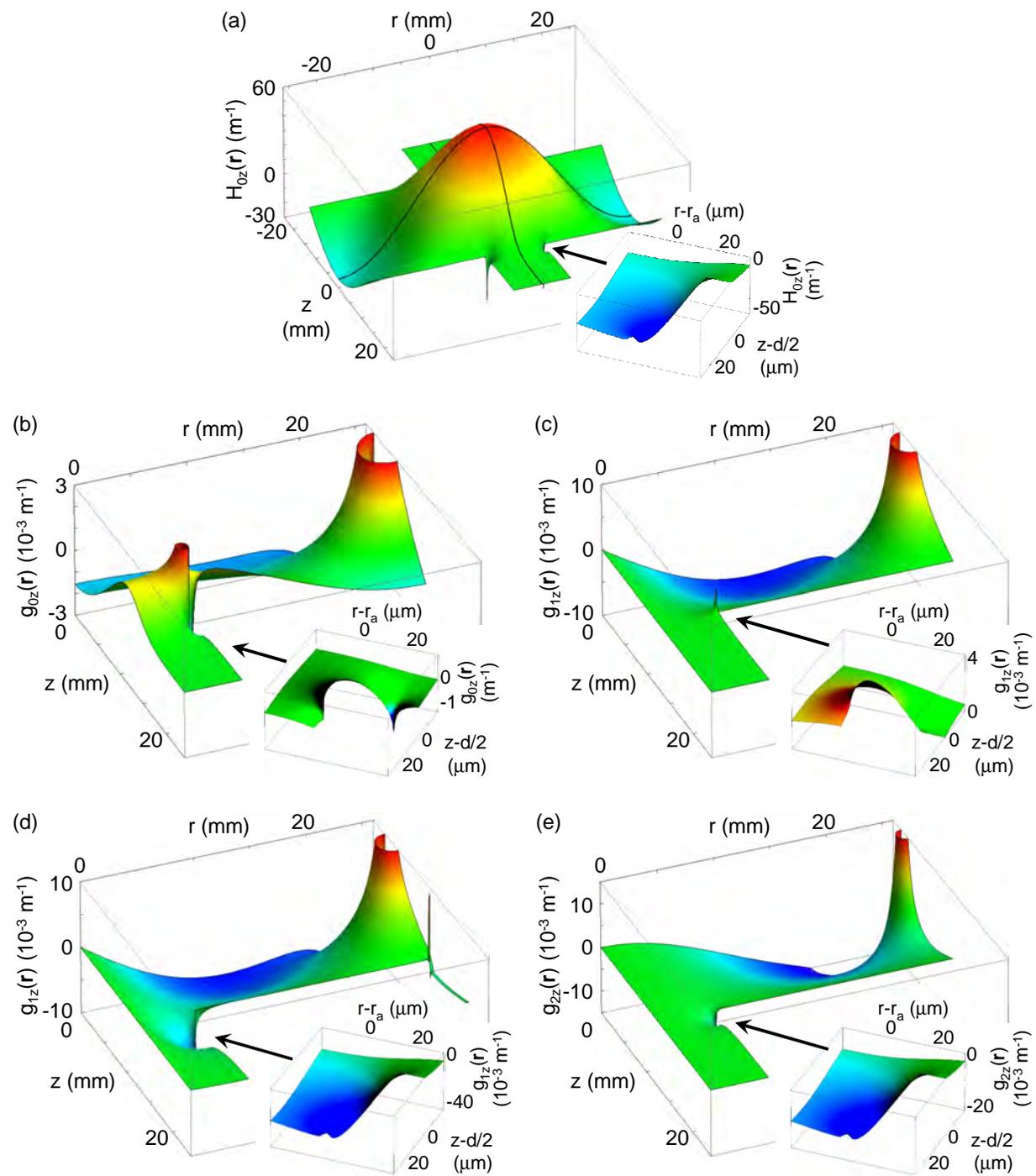

Li Fig 4

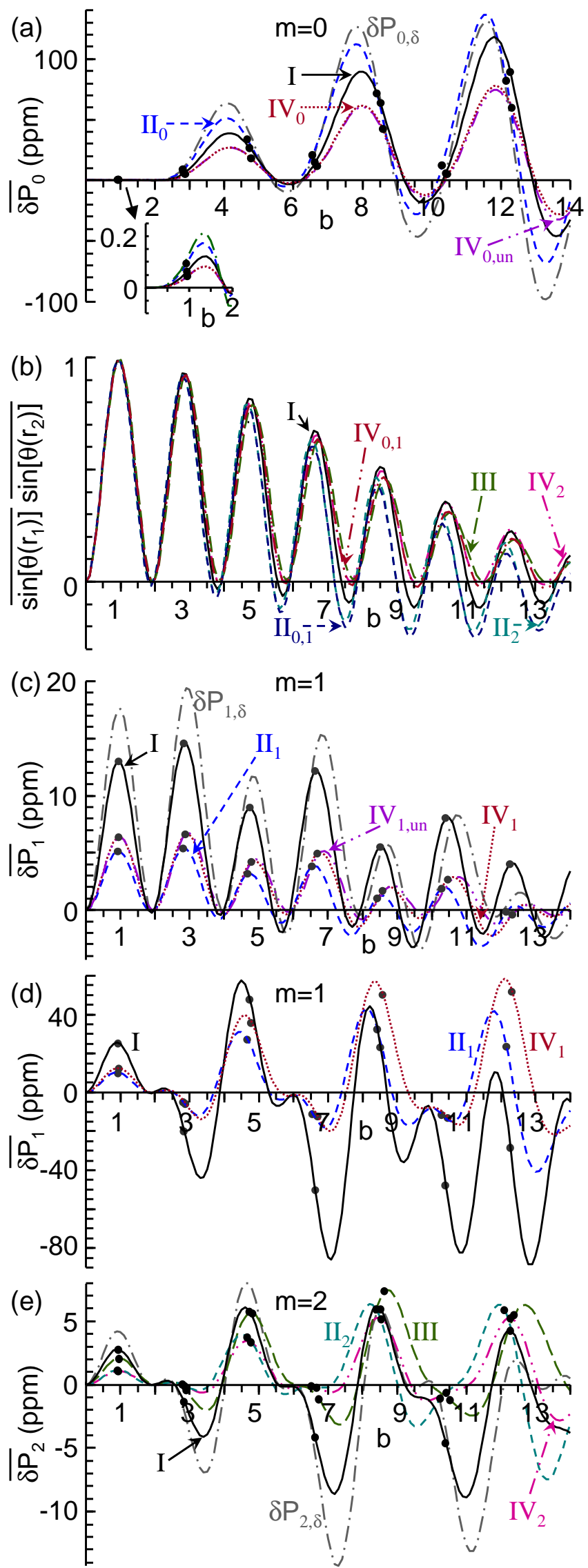

Li Fig 5

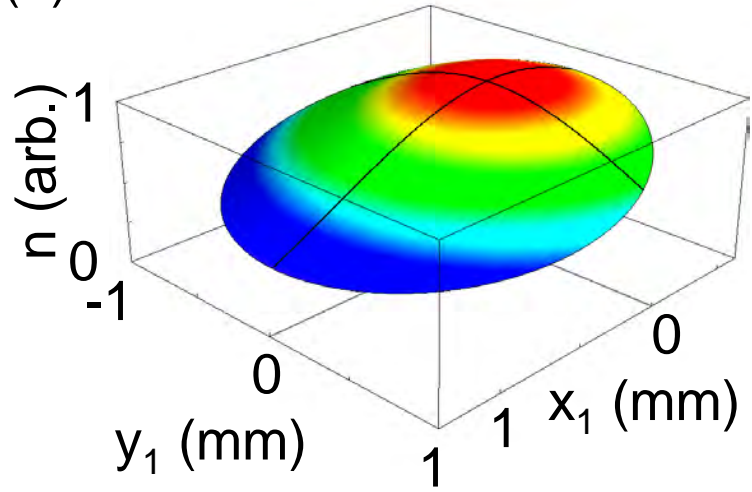 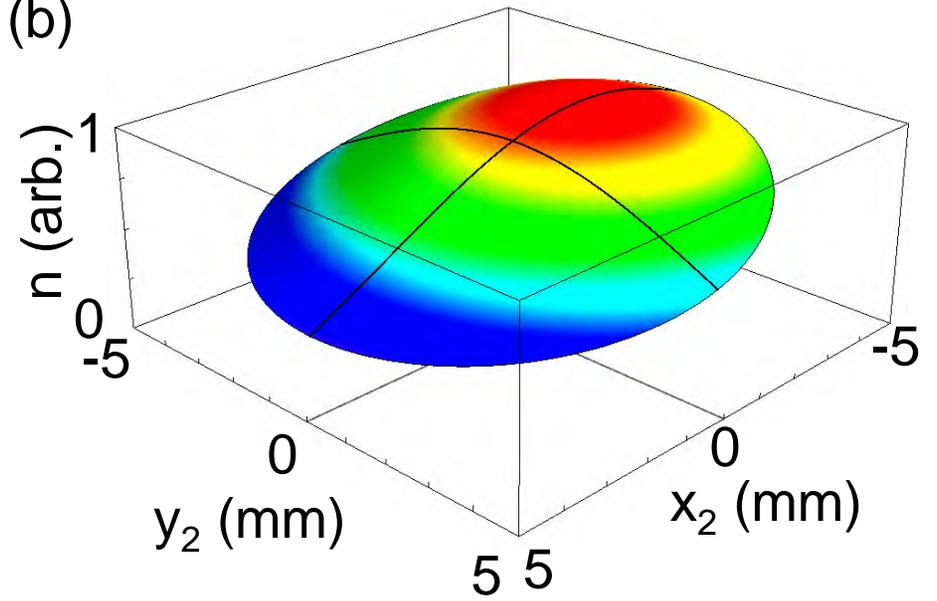

Li Fig 6

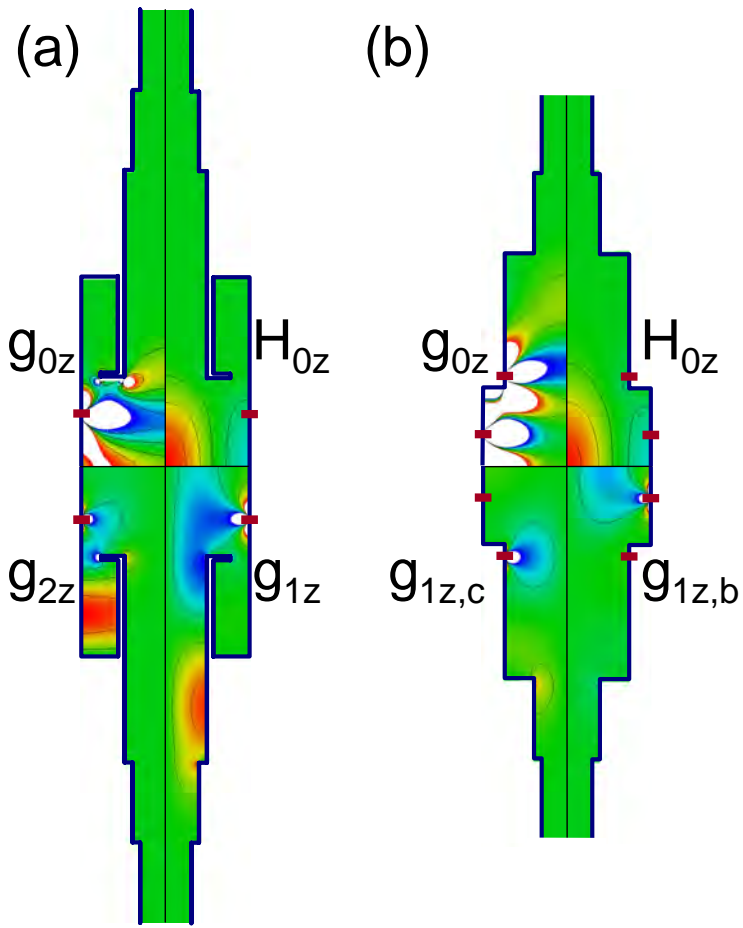

| Dimensions | (a) Radial Width | (a) Length | (b) Radius | (b) Length |
|---|---|---|---|---|
| Body | 21.28 | 44.59 | 21.28 | 38.97 |
| m=0 Cutoff | 10.5 | 51.7 | 15.75 | 33.6 |
| m=1 Cutoff | 8.4 | 20 | 8.4 | 20 |
| Final Cutoff | 6.5 | >20 | 6.5 | >20 |
| Mode filter | 9.28 | 24.16 | | |
| Filter Coupling | 5 | 1 | | All in (mm) |

The aperture for atoms has a radius $r_a$ = 6mm (not shown).

| Feed heights | | ±13.22 | | ±7.9 & ±22.48 |
|---|---|---|---|---|

Li Fig 7

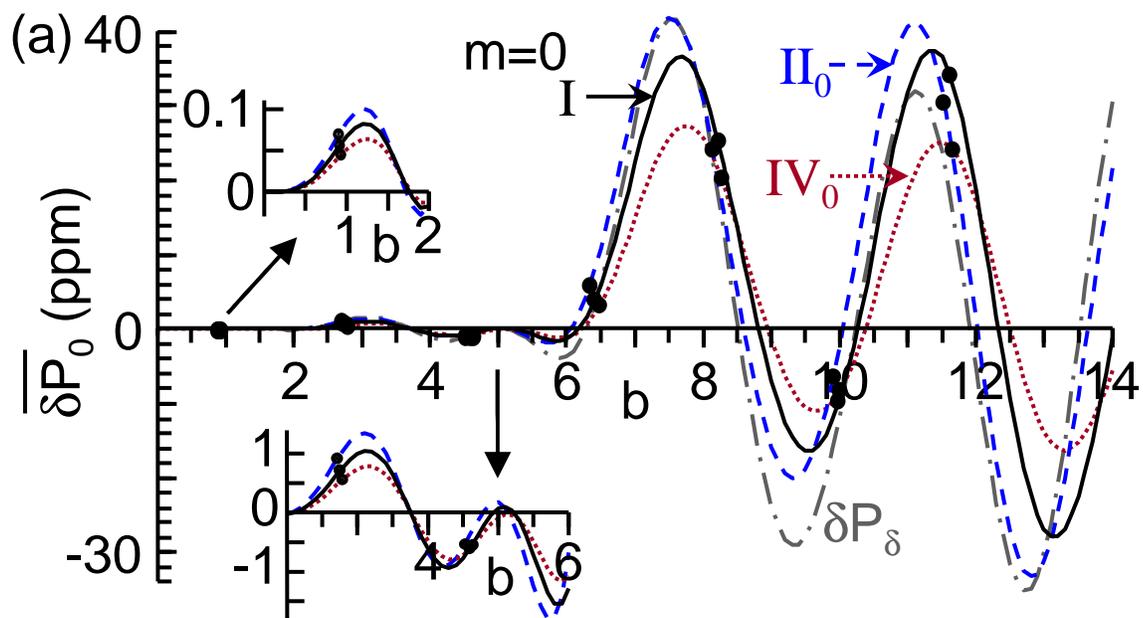
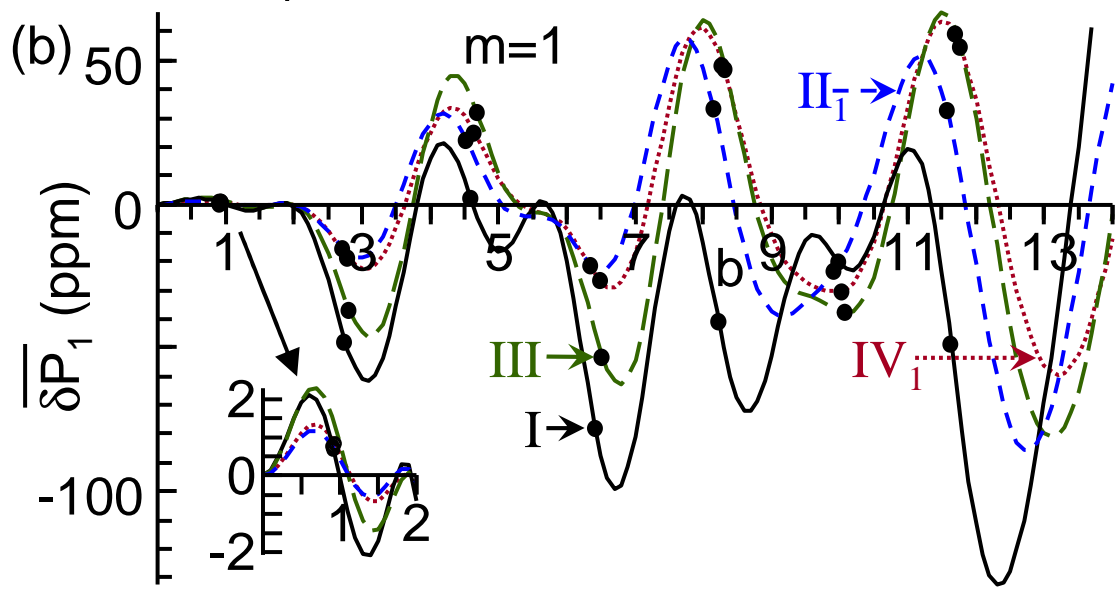
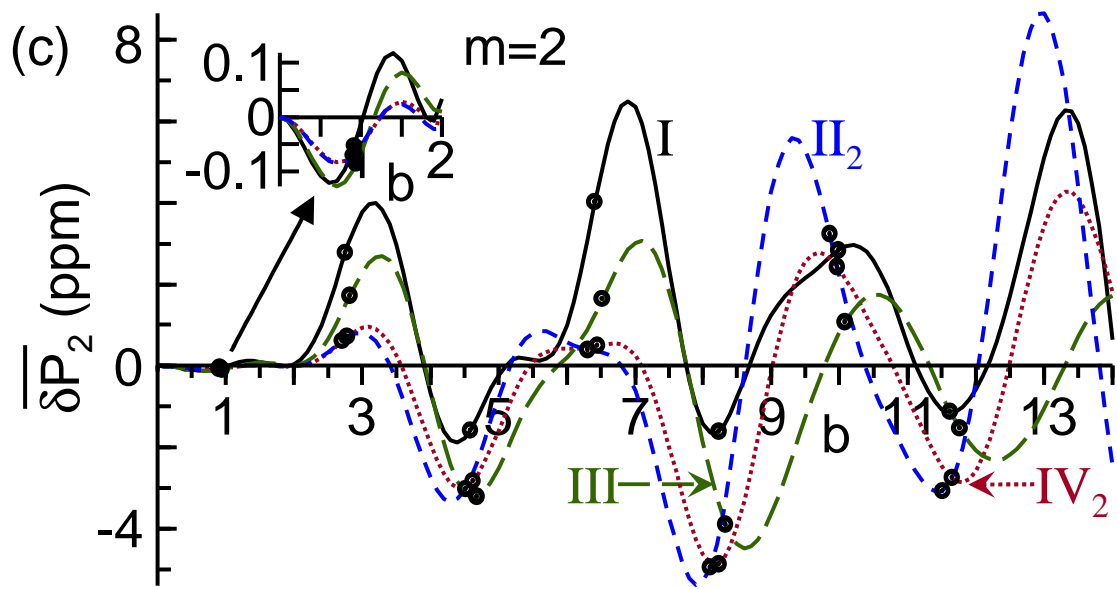

Li Fig 8

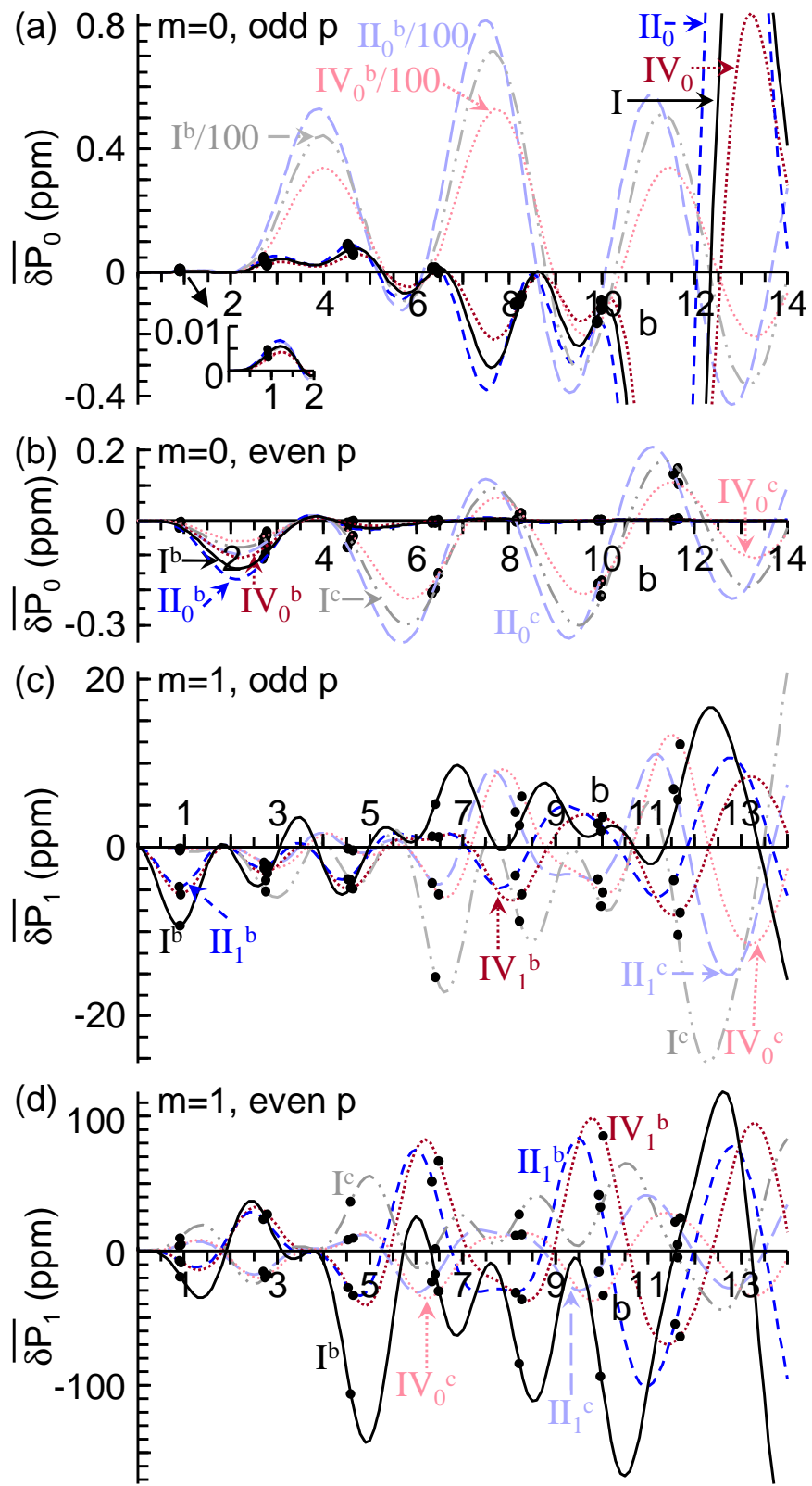

Li Fig 9